\pdfoutput=1

\documentclass[12pt]{article}
\usepackage{eurosym}
\usepackage[T1]{fontenc}
\usepackage[utf8]{inputenc}
\usepackage{geometry}
\usepackage[active]{srcltx}
\usepackage{color}
\usepackage{bbm}
\usepackage{amsthm}
\usepackage{amsmath}
\usepackage{booktabs}
\usepackage{setspace}
\usepackage{lscape}
\usepackage[authoryear]{natbib}
\usepackage[font=footnotesize, width=1\textwidth]{caption}
\usepackage[T1]{fontenc}
\usepackage{graphicx}
\usepackage{times}
\makeatletter
\usepackage{amsfonts}\usepackage[page]{appendix}
\usepackage{color}
\usepackage[usenames,dvipsnames,svgnames,table]{xcolor}
\usepackage{pgfplots}

\usepgfplotslibrary{dateplot}
\pgfplotsset{compat=1.5}

\newtheorem{proposition}{Proposition}

\pdfoutput=1

\begin{document}

\title{Is Factor Momentum More than Stock Momentum?
\thanks{\scriptsize We are especially grateful to Jean-Philippe Bouchaud for his feedback. We also thank Mark Potters, Philip Seager, as well as CFM seminar participants for useful comments.}
\bigskip
} 
\vspace{.5cm}
\author{Antoine Falck\thanks{Capital Fund Management International Inc., The Chrysler Building -- 55th floor, 405 Lexington Avenue, New York, NY, 10174}
\and
Adam Rej\footnotemark[2]
\and David Thesmar \footnotemark[2] \thanks{MIT, NBER \& CEPR}
}
\maketitle

\begin{abstract}
Yes, but only at short lags. In this paper we investigate the relationship between factor momentum and stock momentum. Using a sample of 72 factors documented in the literature, we first replicate earlier findings that factor momentum exists and works both directionally and cross-sectionally. 
We then ask if factor momentum is spanned by stock momentum. A simple spanning test reveals that after controlling for stock momentum and factor exposure, statistically significant Sharpe ratios only belong to implementations which include the last month of returns. 
We conclude this study with a simple theoretical model that captures these forces: (1) there is stock-level mean reversion at short lags and momentum
at longer lags, (2) there is stock and factor momentum at all lags and (3) there is natural comovement between the PNLs of stock and factor momentums at all horizons.
\end{abstract}

\newpage

\section{Introduction}

This paper analyzes factor momentum. In recent years, a body of literature in asset pricing has analyzed the properties
of a large number of risk factors (\cite{...and}, \cite{pontiff}, \cite{HXZ}, \cite{GFX}). These factors are constructed
 by sorting stocks along various characteristics and are evidence that stock returns are predictable and co-move. As it
 turns out, recent research suggests that factor returns are themselves predictable (\cite{cohen},\cite{valentin}) and
 in particular exhibit momentum \citep{M2}: A factor that was performing well tends to perform well in the future. At
 the same time, it has been known for long that stocks also exhibit momentum (\cite{jegadeesh1993}). Factor and stock
 momentum are likely to be interconnected: Since stocks are exposed to factors, and factors have persistent returns,
 factor momentum may explain stock momentum. Conversely, since factors are portfolio of stocks, stock momentum may
 mechanically lead to factor momentum. This paper investigates the relation between these two momentums. We find that
 factor momentum is difficult to distinguish from stock momentum: The only difference is at monthly time scale, where
 stocks mean revert while factors exhibit strong momentum.

We shall proceed in three steps. First, we use US data and build a dataset that contains 72 documented signals
published in the literature. We replicate the construction of factors as long-short dollar-neutral portfolios and
further hedge their residual market exposure dynamically. An equal-risk average of these factors generates a Sharpe
ratio of 0.96, thereby confirming the soundness of our replication exercise. Consistent with many investment strategies
becoming crowded over time, we find that average risk-adjusted performance tends to taper off in the late 2000s. We then
 document that these factors exhibit momentum. As \cite{M2}, we explore both types of factor momentum: cross-sectional
 (buying the most successful factors and selling the least performing ones) and directional (buying factors with past
 positive returns and selling factors with past negative returns). We find that both strategies deliver significant and
 similar risk-adjusted performance, with a Sharpe ratio around 1. We then analyze ``mechanical'' drivers of factor momentum. Since factor momentum buys the best performing factors and sells the unperforming ones, factor momentum contains a mechanical exposure to the mean and spread of factor returns, a phenomenon shown in \cite{lo}. Controlling for mean factor return, we still find that factor momentum is present, in line with results shown by \cite{M2} using a different set of factors and a slightly different spanning approach. The bottom line is that factor momentum is a resilient anomaly that begs for an explanation.

In the second step of our analysis, we analyze the distinction between factor and stock momentum. We use the following
approach. In the spirit of \cite{jegadeesh1993}'s analysis of stock momentum, we analyze a full range of definitions of
factor momentum using various lags (how many most recent months of data one discards when constructing the signal) and
various ``holding
periods'' (number of months worth of stock returns used to compute momentum). We first find that factor momentum is
present for a large range of lags and holding periods. We then compute the excess performance of each one of these
momentums after controlling for factor exposure. Consistently with our first step, we find that controlling for mean
factor exposure removes about 0.2 points of Sharpe from all factor momentums (i.e., irrespective of lag and holding
period). Controlling for stock momentum further reduces the Sharpe of nearly all factor strategies, except for the
smallest possible lag (one month). Further, the excess return remains particularly strong when one focuses on one-month
lag and one-month momentum time scale, i.e., the monthly mean reversion. Additional analysis confirms that factor
momentum is ``spanned'' by stock momentum and factor exposure, except at one-month time scale. Put differently, factor
returns are persistent at the monthly time scales, while stock returns mean revert. Otherwise, factor momentum is not
distinguishable from stock momentum.

In our third step we develop a simple model designed to encapsulate these findings. The key question is whether factor
momentum can co-exist with stock mean reversion at short horizons and stock momentum at longer horizons. In our model
there are three types of investors: noise traders, positive feedback traders who buy stocks exposed to an arbitrary
factor when it performs well and contrarian traders, who buy stocks whose price is high. The model captures the three
features present in the data: (1) there is stock-level mean reversion at short lags and momentum at longer lags, (2)
there is factor momentum at all lags and (3) there is natural comovement between the PNLs of stock and factor momentums at all horizons (short- and long-term).

Our paper contributes to the literature on factor timing. This literature looks for evidence that risk factor returns
are themselves predictable. Most papers in this space focus on mean-reversion in factor returns. \cite{cohen}, and more
recently \cite{valentin} build on the idea that factors are measures of risk premia, which tend to be mean-reverting
\citep{CS}: A factor is expensive when stocks in its long (short) leg have a high (low) market to book ratio. Both
papers find evidence consistent with the idea that expensive factors tend to have lower future returns. \cite{cohen}
focuses on the value premium, while \cite{valentin} focus on the first PCs of a large group of factors. In the same spirit, \cite{hanson} find that factors that short stocks that are otherwise heavily shorted tend to mean revert.

Our paper focuses instead on shorter-term persistence in factor returns. We follow \cite{M2} and provide further
evidence of factor momentum using a different set of factors (we have 72 factors while they have 20 in their baseline
study; moreover our factors are defined in US only) and a slightly different portfolio construction. Like these authors,
 we find that directional factor momentum is marginally stronger than the cross-sectional implementation, with both
 strategies generating most of their returns from their long legs. But the main difference between their study and ours
 is in the test
 differentiating stock and factor momentum. They construct factor momentum as a strategy based on the past 12 months of
 returns, while stock momentum is also based on the past 12 months of returns \emph{excluding} the last month. They find
  that factor momentum is not spanned by stock momentum. Our strategy consists of looking at different definitions of
  factor momentum, where we vary both the lag and the holding period. This allows us to uncover the source of the
  discrepancy between the two momentums: At one-month time scales, stocks mean-revert while factors have persistent
  returns. If we
  exclude the last month of returns, factor momentum is subsumed by stock momentum. We provide a model consistent with our findings.

The paper is structured as follows. Section \ref{sec:data} describes the data and the construction of our factor zoo.
Section \ref{step1} provides evidence of factor momentum and tests for mechanical drivers of momentum. Section
\ref{step2} decomposes factor momentum into various lags and horizons. This decomposition shows the key role of the lag
in differentiating between both types of momentum. Finally, Section \ref{step3} proposes a model consistent with
stylized facts that we uncovered. Section \ref{conclu} concludes.

\section{Data \& Factor Construction}
\label{sec:data}

We construct time series of returns for 72 risk factors documented in the literature (see list in Appendix \ref{sec:list_factors}). These risk factors use data from a CRSP-COMPUSTAT merged sample, so they only use accounting and price information. This sample runs from January 1963 to April 2014. In our merger of CRSP and COMPUSTAT, we make sure that there is no look ahead bias in financial statement availability. We assume that financial statements for the last fiscal year become available four months later. This is conservative as earnings are typically anounced between 2 and 3 months after the end of the fiscal year. But it ensures that there is little look ahead bias in our data.

While our implementation of the 72 sorting variables is faithful to the papers we replicate, we propose certain
improvements to risk management in order to be closer to what is common practice in asset
 management. First, we restrict our analysis to the 1,000 most liquid stocks on CRSP. This is in order to ensure that
 the factors trade liquid stocks and have reasonable capacity. Second, we construct and risk manage each factor using
 the following methodology. For each factor in Appendix \ref{sec:list_factors}, we first compute daily factor returns by
  implementing the characteristic proposed in the original factor paper and using the portfolio construction method
  proposed by the authors. We resample the PnLs monthly. Next, we remove any residual exposure to being long the market
  by beta-hedging, i.e. we run a rolling 36 months regression of the factor PnL onto market returns to determine the
  (past) beta and use this beta for the following month to debias PnLs. Finally, we risk manage the resulting PnL by
  dividing it by the (rolling) standard deviation of returns computed over past 36 months and lagged one month. This
  ensures
  that each beta-hedged factor PnL has a relatively constant volatility and is thus closer to how factor investing would be implemented in practice. We report the distribution of Sharpe ratios and t-stats of our 72 factors in Table \ref{tab:stats_zoo}. As it becomes apparent, a typical factor Sharpe ratio is low (0.27), suggesting potential overfitting or arbitrage, an issue we explore in a companion paper.

\bigskip

\begin{table}[tbph]
\caption{Summary Statistics on our Sample of 72 Factors}
\label{tab:stats_zoo}
  \begin{center}
\begin{tabular}{lrrrr}
\toprule
 &    Mean & Median & p25 & p75 \\
\midrule
\\
Annualized Sharpe       &  0.27 &     0.35 &         -0.02 &      0.55 \\
\\
t stat & 1.73 &     2.11 &         -1.25 &      3.64 \\
\bottomrule
\end{tabular}

  \end{center}
  \footnotesize{Note: Annualised Sharpe and t stats, computed with monthly returns. All factor are beta-neutralised and risk-managed,
  computed on CRSP 1,000 most liquid stocks from 1963 to 2014.}
\end{table}

\bigskip

\begin{figure}[tbph!]
  \caption{Mean Factor Return}
  \label{fig:mean_factor_return}  
  \begin{center}
  \includegraphics[width=12cm]{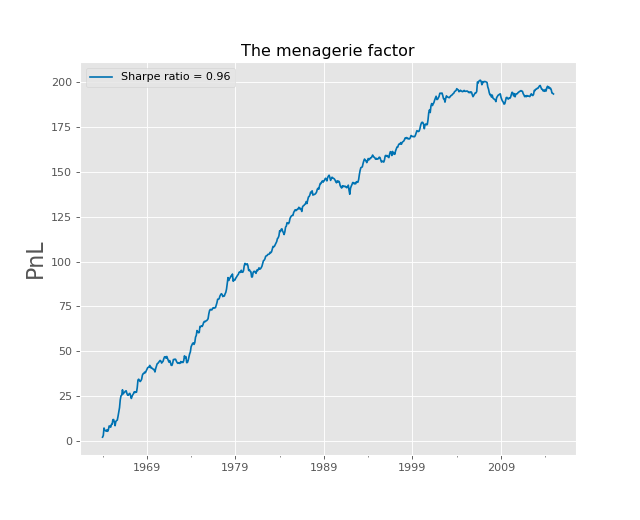}
  \end{center}
  \footnotesize{We report here the cumulative return of factor risk parity with 72 factors. The investment strategy is beta-hedged in order to filter out exposure to the market and run at a constant risk target. }
\end{figure}

We report in Figure \ref{fig:mean_factor_return} the cumulative return of factor risk-parity. This investment strategy consists in being long an equal amount of risk of each of the 72 factors and is run at a constant overall risk target. In what follows, we will refer to this strategy per ``menagerie'' (all of the ``animals'' in our ``zoo'' of factors). The Sharpe ratio of this strategy over the entire period is 0.96. The cumulative return does however show signs of tapering off starting in the mid 2000s. This is consistent with the idea, described in \cite{pontiff} among others, that after publication the strategies listed in these papers perform less well. 

\bigskip

\section{Evidence on Factor Momentum}
\label{step1}

\subsection{Defining momentum}
\label{definition}

Let $F_{t}$ be a vector of monthly factor returns, with returns computed between months $t-1$ and $t$ (and each factor
constructed using information available at $t-1$). Each component of this vector corresponds to one of our risk-managed
factors presented in Section \ref{sec:data}, i.e. they are dynamically hedged with respect to market exposure and scaled
 by rolling volatility. $N$ is the number of components. We first define the ``menagerie'' as investing \$1 in each one of these factors (so the amount of risk in each factor is the same). Hence the PNL of the menagerie at the end of month $t$ is:

$$\pi^{me}_t =e'F_{t}$$

\noindent where $e=(1,1..1)$ is a vector of ones. We denote the transpose by $'$.

\bigskip

We then define cross-sectional and directional factor momentum as in \cite{M2}. The key difference is that we sort factors based on past returns computed from $t-m-n$ to $t-m$. Put differently, $m$ is the most recent price used (the ``lag'') and $n$ is the number of months used (the ``holding period''). To make things clear, we note:

 $$F_t(m,n)=\sum_{k=m}^{k=m+n-1}{F_{t-k}}$$ 

\noindent the cumulative return between $t-m-n$ and $t-m$. Note that as a special case $F_t(0,1)=F_t$.

The PNL of cross-sectional factor momentum is given by:
\begin{equation}
\label{eq:fact_mom}
\pi^{XS}_{t}(m,n) =\text{rank}\left(F_{t}(m,n)\right)'F_{t}
\end{equation}

\noindent where the $\text{rank}$ operator transforms a vector with $N$ components into a vector of ranks with $N$ different, equally-spaced components in the range $\langle -1, 1 \rangle$. By construction the cross-sectional momentum is always dollar neutral, i.e. the sum of weights is equal to zero. 

The PNL of directional momentum does not impose dollar neutrality. We define it as:
\begin{equation}
\label{depardieu}
\pi^{TS}_{t}(m,n)\ =\text{sgn}\left(F_{t}(m,n)\right)'F_{t}
\end{equation}

\noindent For instance, if all factors have performed positively in the past period, the factor momentum portfolio will
(temporarily) be the same as the menagerie. The net leverage of this strategy varies over time: it increases (decreases)
 when factors have positive (negative) returns.

The above definitions (menagerie, XS and TS momentum) are that of ``raw'' portfolios, in the sense that they are not
risk managed. The empirical results we present below are for the ``risk-managed'' versions of these PNLs, which are
obtained by normalizing the returns with their rolling 36 month volatility lagged one month. Remember that each one of
the factors is itself already risk-managed.

Last, for clarity of exposition, we focus in the sections that follow on directional momentum, but our results with cross-sectional momentum are very similar. We report them in Appendix \ref{sec:appendix_xs}.

\subsection{Factor Momentum: First Evidence}

Here, we redo the analysis in \cite{M2} using our own set of factors and report results in Table \ref{tab:stats_mom}.
The
first line shows the performance of the menagerie strategy, which has a Sharpe ratio of 0.96 (as we showed in Figure \ref{fig:mean_factor_return}) and a t-stat of 6.86.

\begin{table}[tbph]
\caption{Momentum in Our Set of Factors}
\label{tab:stats_mom}
  \begin{center}
\begin{tabular}{lrr}
\toprule
{}  & Sharpe ratio & $t$-value \\
\midrule
Menagerie        &         0.96 &      6.86 \\
TS               &         1.16 &      8.22 \\
\ \ TS, winners  &         1.29 &      9.11 \\
\ \ TS, losers  &         0.43 &      3.03 \\
XS               &         1.01 &      7.12 \\
\ \ XS, winners  &         1.20 &      8.52 \\
\ \ XS, losers  &         0.46 &      3.24 \\
\bottomrule
\end{tabular}
  \end{center}
  \footnotesize{Note: Annualised statistics computed with monthly returns. We compute factor momentum signals using $n=10$ and $m=2$.
  Factors are beta-neutralised and risk-managed,
  computed on CRSP 1,000 most liquid stocks from 1963 to 2014.}
\end{table}

The next lines of Table \ref{tab:stats_mom} show the performance of factor momentum. We calculate the direct returns of cross-sectional
 and directional momentum (using formulas (\ref{eq:fact_mom}) and (\ref{depardieu})). We set $n=11$ and $m=2$ for now (we come back to varying $m$ and $n$ later).
Time series and cross-sectional momentum both have high Sharpe ratios and the directional implementation outperforms the cross-sectional one by $0.15$ in Sharpe. \cite{M2} also find that directional factor momentum outperforms, but they find a somewhat larger difference in Sharpe .26 (Their Table 2). There are three main differences between their methodology and ours. First, we are using a larger set of factors. Second, we are constructing our factors using the 1,000 largest U.S. stocks only. Our construction thus excludes illiquid microcaps. Last, we risk-manage and beta-neutralize factors' PNLs. In spite of these significant differences, we manage to reproduce their headline results quite closely.

Another salient feature of Table \ref{tab:stats_mom} is that most of the return on factor momentum is borne by the long
leg of the strategy (the ``winners''). \cite{M2} find similar results. One possibility is that this whole effect is
driven by somewhat unsophisticated, long-biased investors piling into factors that performed well, as in \cite{hanson}).
 We rely on this idea in our model of Section \ref{step3}.

\subsection{Accounting for mechanical exposure of momentum to factors}
\label{mechanical}

The key role of the long leg of factor momentum also suggests that part of the performance may simply come from outright exposure to factors. This concern is particularly relevant for directional momentum. This intuition can easily be confirmed in a momentum decomposition \`a la \cite{lo}. Let us focus on one single factor $f_t$ (an element of $F_t$). Assume that this factor follows an AR1 process $f_t=(1-\rho)\mu+\rho f_{t-1}+u_t$, where $\mu$ is the long-run excess return of the factor. Then, the conditional and unconditional PNLs of directional momentum on this factor alone are given by:

\begin{align*}
E_{t-1} (f_{t-1}f_t)&=\rho f_{t-1}^2 + (1-\rho)\mu f_{t-1}  \\
E (f_{t-1}f_t)&=\rho \sigma_f^2 + \mu^2 
\end{align*}

\noindent These two formulas illustrate why factor momentum is more likely to have a good PNL if the expected return of
the factor is positive $\mu>0$, independently of the strength of factor momentum itself ($\rho$). To fix ideas, set
$\rho=0$. The first equation shows the conditional PNL of the factor momentum strategy, which is non-zero. Indeed when
the factor has performed well ($f_{t-1}>0$), the momentum trader will purchase it and will thus cash in the expected
premium $\mu$: expected profit will be $\mu f_{t-1}>0$, even though factor returns have no real momentum. Another way to
 see this is by looking at the unconditional PNL in the second equation. Even if $\rho=0$, and as long as
 $\mu\neq0$, factor momentum always generates a positive PNL. On average, factors with positive premia are more likely
 to experience two consecutive dates of positive returns. All the same, factors with negative premia are more likely to underperform on two consecutive dates. This ensures $E (f_{t-1}f_t)>0$ even if there is no actual persistence in returns.

\begin{figure}[tbph]
\caption{PNLs of Menagerie and Factor Momentums}
  \label{fig:pnl_mom}
  \begin{center}
  \includegraphics[width=12cm]{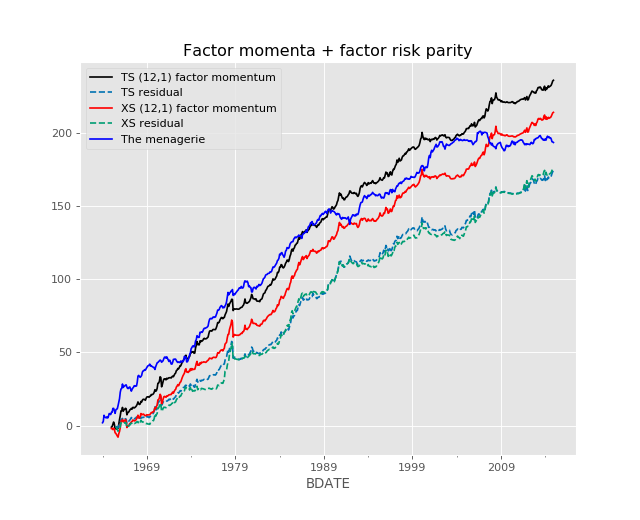}
  \end{center}
  \footnotesize{Cumulative sum of returns to two factor momentum strategies and
  the menagerie factor. All PnLs are rescaled (in a causal way) to target 1\%
  monthly volatility.
  Factors are beta-neutralised and risk-managed,
  computed on CRSP 1,000 most liquid stocks from 1963 to 2014.}
\end{figure}

To account for this mechanical relationship (and isolate the pure effect of factor momentum), we look at the residual of the regression of factor momentum PNL on the menagerie PNL. The regression is run over the entire period as we are not looking to hedge factor momentum, but rather to measure its \emph{de facto} exposure to the menagerie. We show the effect of this control in Figure \ref{fig:pnl_mom} where we report the momentum returns alongside with the corresponding residuals. We can see that the cumulative PNLs of residuals are about 20\% less than that of pure factor momentum. Thus, while the mechanical exposure of momentum to factor returns in non-negligible, factor momentum remains a strong force.

\section{Factor vs Stock Momentum}
\label{step2}

We now turn to the analysis of the link between stock momentum and factor momentum. We start with an analysis of the correlation between the two factors. We then break the analysis down into various lags and holding periods. In this section we use the directional implementation of factor momentum. The results for cross-sectional momentum are similar and are reported in Appendix \ref{sec:appendix_xs}.

\subsection{Correlation with Stock Momentum assuming standard lag and holding period}
\label{firstpass}

To investigate this, a natural object to look at is the correlation between stock and factor momentum. Let us first investigate this correlation analytically. To fix ideas, let us assume that the vector of stock returns $r_t$ (between period $t-1$ and $t$) follows a single factor structure:

$$r_t = \beta f_t + e_t$$

\noindent where $\beta$ is a vector of factor exposures of individual stocks and $e_t$ is the idiosyncratic shock, assumed to be stationary, i.i.d. and most importantly independent of $f_t$ for all lags. The factor can be persistent but we assume it to be homoskedastic -- its conditional variance does not vary with time. We define here the PNLs of (directional) factor and stock momentum strategies as:

\begin{align*}
\pi_t^{\text{F}}&= \underline{f_{t}}' f_{t} \\
\pi_t^{\text{S}}&= \underline{r_{t}}' r_{t}
\end{align*}

\noindent where the lower bars stand for lagged cumulative returns with arbitrary lag $m$ and holding period $n$. In
this section we do not risk manage PNLs in order to obtain closed-form formulas (in the empirical analysis we will stick with the empirical definitions).

In this setting one can show (See Appendix \ref{proof: covariance}) that the covariance between the two PNLs is given by:

\begin{align}
\label{cov equation}
\text{cov} \left(\pi^F_{t+1},\pi^S_{t+1}\right) = \beta'\beta\text{var}\left(\underline{f}_t f_{t+1}\right) = \beta'\beta \text{var} \left(\pi^F_{t+1}\right) > 0 
\end{align}

\noindent which is generally positive, whether factor momentum has a positive PNL or not. The intuition is simple:
Buying winning stocks is equivalent to exposing oneself to the factor. So whenever stock momentum performs well, it is
likely that the underlying factor has performed well. In this very simple setup, all of stock momentum comes from factor
 momentum (as we assumed $e_t$ to be i.i.d.). So even if factor momentum is not present, there will be instances when
 it
 works by accident and consequently stock momentum will work too. The bottom line of this analysis is that the
 covariance between both strategies is mechanically positive, even when none of them has a positive Sharpe ratio. This
 due to the fact that $\beta$ is stable.

\bigskip

Let us now move to the empirical analysis and compute the correlation between the two types of momentum. For factor momentum, we choose $m=1$ and $n=12$. For stock momentum, we use UMD, the stock momentum factor available from the Fama-French data library. UMD sets $m=2$ and $n=11$ in order to exclude short-term mean-reversion. We report this correlation and other statistics in Table \ref{tab:corr_mom}. We use monthly returns. As expected from our algebra, the correlation of factor momentum with UMD is high, around 60\%. This is much higher than the correlation between UMD and the menagerie (0.04\%). Note that this high correlation is thus unaffected whether we look at directional, cross-sectional or residual momentum. The analysis above has shown why part of this correlation is likely to be mechanical.

\begin{table}[tbph]
  \caption{Correlation with the momentum (UMD) factor.}
  \label{tab:corr_mom}
  	\begin{center}
\begin{tabular}{lrrrrrr}
\toprule
{} &  Menagerie &  Menagerie w/o UMD &    TS &  TS residual &    XS &  XS residual \\
\midrule
Correlation &       0.04 &  -0.08 & 0.57 &        0.59 & 0.58 &        0.58 \\
\bottomrule
\end{tabular}

	\end{center}
  \footnotesize{Factors are beta-neutralised, risk-managed and computed on CRSP 1,000 most liquid stocks from 1963 to 2014. UMD sorts stocks based on the last 12 months of returns except for the most recent one ($m=2$ and $n=11$)}. XS and TS factor momentum sorts factors using the last 12 months of returns ($m=1$ and $n=12$).
\end{table}

This very strong correlation between the two types of momentum begs for a ``spanning test''. To do this we regress the
PNL of factor momentum on the PNL of the stock momentum strategy using the entire sample. We then compute the Sharpe
ratio of the residual strategy. Using the above definitions of stock and factor momentum we find a Sharpe ratio of 0.59,
 which is statistically significant. We explore this in greater detail in Section \ref{sec: residual}. The fact that
 stock momentum does not entirely span factor momentum is consistent with \cite{M2} who implement similar spanning tests
 . This analysis is, however, incomplete as (1) it does not take into account mechanical exposure to the menagerie (as
 shown in Section \ref{mechanical}) and (2) the lags $m$ are different for stock and factor momentum. Let us now investigate these two aspects.

\subsection{Performance of stock and factor momentum for different lags and holding periods}
\label{factor_correl}

In this section we vary the holding period $n$ and the lag $m$ and control for their values when analyzing the correlation between the two types of momentum. It is a priori important to control for these parameters, as they greatly affect the performance of momentum, as it is well known since at least \cite{jegadeesh1993}.

\begin{figure}[tbph!]
  \caption{Sharpe Ratios of Stock and Factor Momentum \\
  as a Function of $m$ and $n$}
  \label{fig:sr_stock_factor_mom_ts}  
  \begin{center}
  \includegraphics[width=10cm]{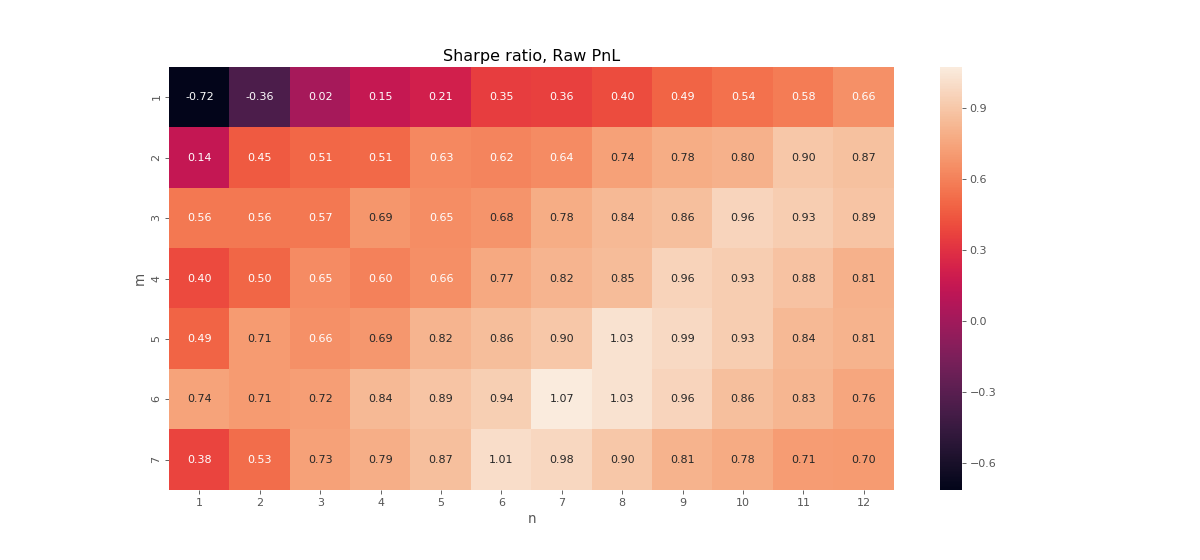}
  \includegraphics[width=10cm]{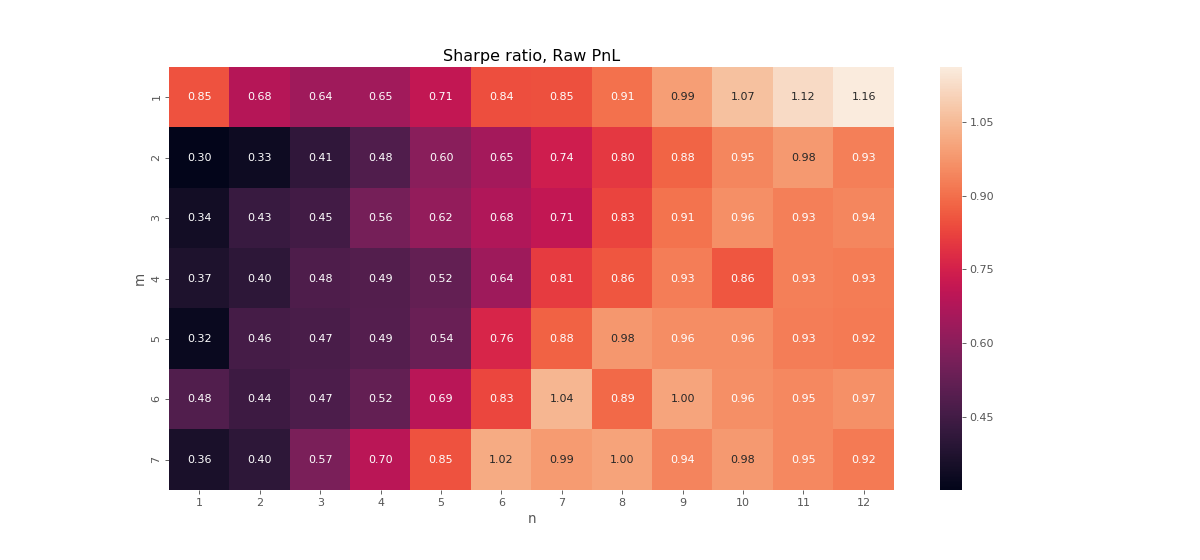}
  \end{center}
  \footnotesize{Top: Sharpe ratios of the cross-sectional stock momentum.
  Stock returns are risk-managed, the universe is CRSP 1,000 most liquid stocks from 1963 to 2014.\\
  Bottom: Sharpe ratio of the directional factor momentum.
  Factors are beta-neutralised, risk-managed and computed on CRSP 1,000 most liquid stocks from 1963 to 2014.}
\end{figure}

In Figure \ref{fig:sr_stock_factor_mom_ts}, before looking at correlations, we first show the performance of both
strategies for different values of $m$ and $n$. Panel A focuses on stock momentum and reproduces three classical results
. First, there is short term reversal for $(m,n) = (1,1)$ as evidenced by the very dark cells (indicating negative Sharpe of momentum) in the upper left corner of the heat map. Second, the classic momentum ($m=2$, $n=10$) has a high Sharpe ratio (0.8). Third, a strongly lagged momentum ($m=6$, $n=6$) has an impressive Sharpe ratio, around 1 as previously noticed by \cite{novymom}. Put differently, there is a northeast-southwest diagonal of light colors on the right side of the heat map, which is an evidence of the strength of ``lagged momentum''. In panel B we focus on factor momentum. Two salient patterns emerge. First, there is \emph{no} short-term reversal in factor returns. Past factor returns \emph{always} predict stronger future returns, even at short time scales. It is clear that having the most recent month of return increases the performance of the strategy: A big fraction of the strength of factor momentum comes from the first lag, \emph{precisely where stock momentum has a negative performance}. Another emerging pattern is that strongly lagged factor momentum, like for stocks, performs quite well. Overall, this analysis highlights the one key difference between stock and factor momentum: in the short-run, stocks exhibit reversals while factors momentum.

\begin{figure}[tbph]
  \caption{Correlation between Stock Momentum and Factor Momentum \\
  For different levels of $m$ and $n$}
  \label{fig:corr_m_n}
  \begin{center}
  \includegraphics[width=15cm]{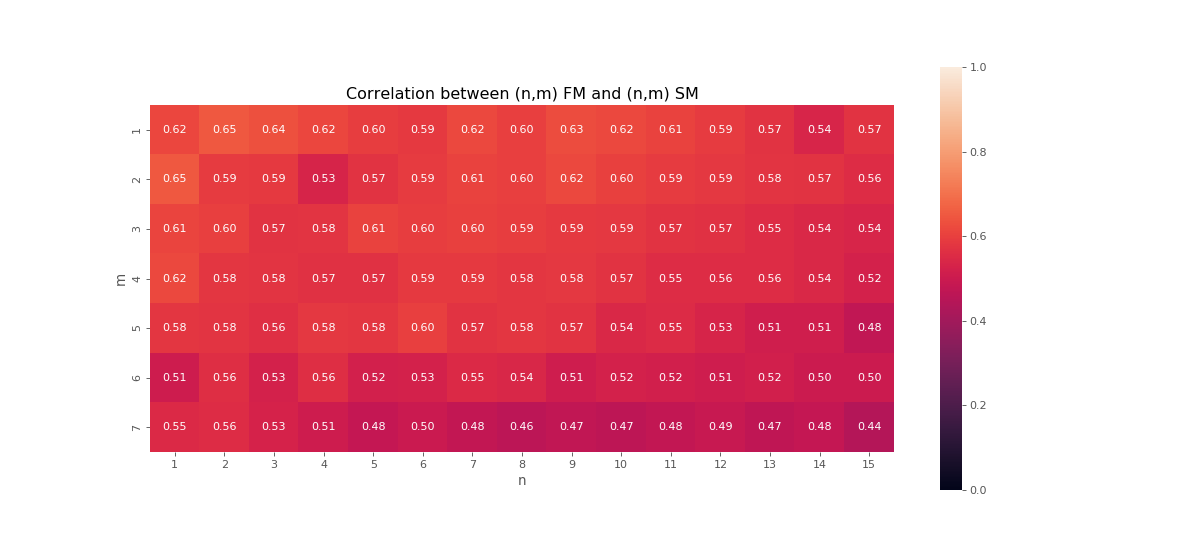}
  \end{center}
  \footnotesize{Stock returns are risk-managed, the universe is CRSP 1,000 most liquid stocks
  from 1963 to 2014.}
\end{figure}

We now move to computing the correlation between the two types of momentum. Results are reported in Figure
\ref{fig:corr_m_n}. The heat map delivers a simple yet impressive result: the correlations are remarkably
stable, around .5 across the board.

The stability of the correlation across values of $(n,m)$ could be a priori surprising since we have just uncovered that
 for small values of $m$ and $n$ stocks and factor returns behave differently. So while it is natural to expect that the
  two strategies are correlated for $m>1$, it is more surprising to find that they are positively correlated when their
  Sharpe ratios have opposite signs. The analysis of Section \ref{firstpass} provides an explanation for this. As shown
  in Equation (\ref{cov equation}) persistent factor exposure of stocks mechanically drives part of the correlation
  between both strategies: when factors perform well two periods in a row, stocks with positive beta do the same, while
  stocks with negative betas consistently sell off. Thus, this correlation is present, and does not depend on how well
  or badly the two momentums perform unconditionally.

\subsection{Residual of factor momentum on stock momentum}
\label{sec: residual}

We now proceed to analyze the residual of factor momentum on stock momentum. We have seen in Table \ref{tab:corr_mom} that factor momentum is not spanned by UMD (the residual has a Sharpe of 0.6). But now we want to incorporate what we have learnt by varying $m$ and $n$, i.e. that the Sharpe of the two momentums vary quite differently as functions of lag and holding period. In addition, we also know from Section \ref{mechanical} that factor momentum is correlated with the menagerie factor, so we also want to include the latter in our spanning test.

\begin{figure}[tbph!]
  \caption{Sharpe Ratio of Residual Factor Momentum \\
	After Taking out Stock Momentum, Menagerie and Market}
  \label{fig:residual}
  \begin{center}
  \includegraphics[width=15cm]{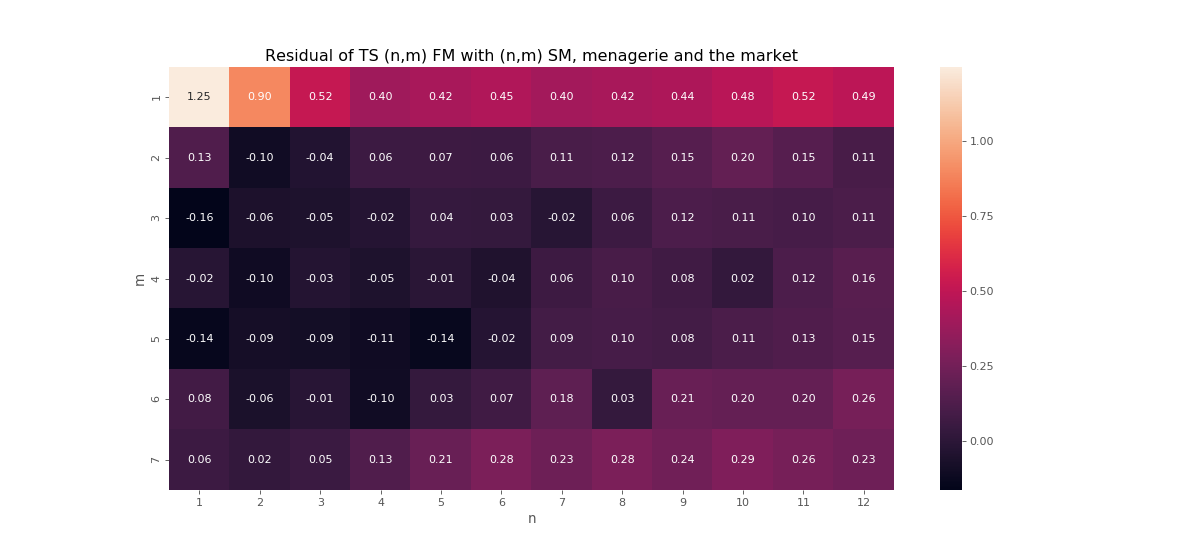}
  \includegraphics[width=15cm]{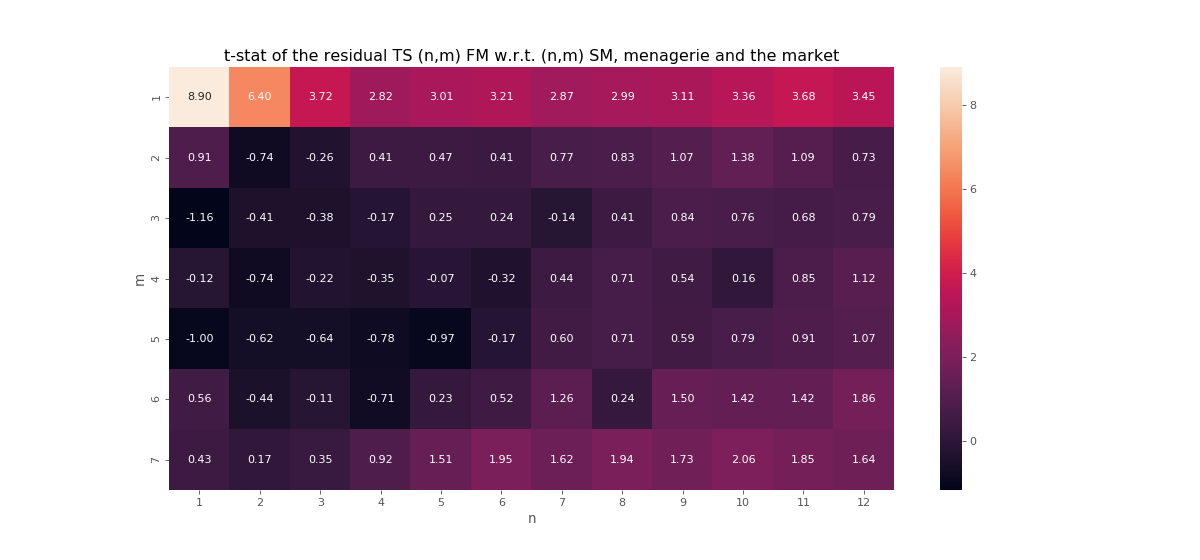}
  \end{center}
  \footnotesize{Stock returns are risk-managed, the universe is CRSP 1,000 most liquid stocks. Sample period
  from 1963 to 2014.}
\end{figure}

Thus, in Figure \ref{fig:residual}, we show the Sharpe ratio of the residual of factor momentum after controlling for stock momentum, the menagerie factor and the market. All regressions use the entire sample (these are spanning tests, not hedging regressions). This residual is computed separately for each value of the lag $m$ and holding period $n$.

The heat map in Figure \ref{fig:residual} suggests that that there is more to factor momentum than stock momentum, but
only at lag $m=1$, i.e. when we include the last month of returns. As soon as $m>1$, the Sharpe ratios of residuals
become insignificantly different from zero. This is consistent with the observation made in the previous section. The
main difference between factor and stock momentum is that for $m=n=1$ stocks mean revert in the cross-section while
factor returns persist. Our next section provides a framework that combines these effects.

\begin{figure}[tbph!]
  \caption{Sharpe Ratio of Residual Stock Momentum \\
	After Taking out Factor Momentum, Menagerie and Market}
  \label{fig:residual2}
  \begin{center}
  \includegraphics[width=15cm]{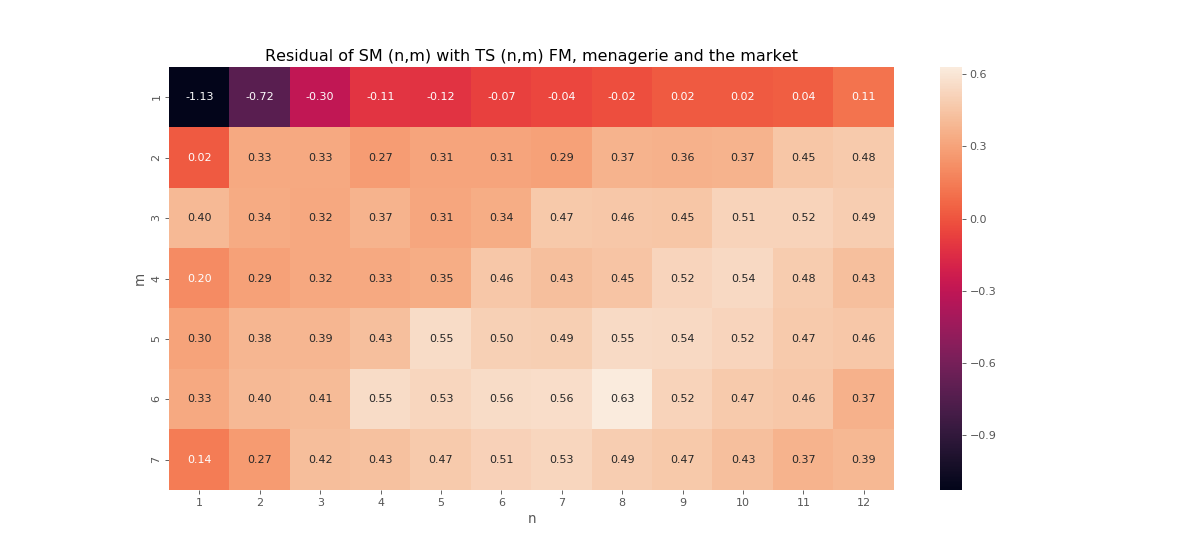}
  \end{center}
  \footnotesize{Stock returns are risk-managed, the universe is CRSP 1,000 most liquid stocks,
  from 1963 to 2014.}
\end{figure}

Before we proceed, however, we would like to know whether stock and factor momentum are identical phenomena or whether factor momentum is just subsumed by stock momentum. In order to do this, we implement a spanning test opposite to the one proposed above, i.e. we regress stock momentum on factor momentum, menagerie and market for different values of $m$ and $n$ in order to single out the effect of lagging the signals. We report the results in Figure \ref{fig:residual2}. For $m=1$, due to mean-reversion, stock momentum is either negative or completely explained by the independent variables. For other values of $m$, however, residual stock momentum is still significant for many cells in the heat map. This suggest that, while factor momentum is just stock momentum when $m>1$, stock momentum is really something else for all values of $m$.

\section{Framework}
\label{step3}

In this section, we will develop a heuristic model that encapsulates the three main stylized facts discussed in the
previous Sections: (1) there is stock-level mean reversion at short lags and momentum at longer lags, (2) there is factor momentum at all lags and (3) there is a natural co-movement between the PNLs of stock and factor momentums at all horizons (short- and long-term).

\subsection{Setup}

We assume that there are $N$ different stocks and that there exists a factor strategy that attracts investors'
interest. This factor strategy is captured by a set of (normalized) positions on each of the $N$ stocks, which we assume
 fixed to simplify the exposition. We denote
them by $w$. We postulate that stock returns satisfy the equation below:

$${r}_{t} = {\mu} + \epsilon_{t} + \left(\textrm{price impact due to flows into/out of factor investing}\right)_t$$

\noindent which can be derived from an equilibrium asset pricing model where three types of investors are present: Noise traders (who generate the mean-reverting noise $\epsilon_t$), factor investors (described below) and contrarian traders who have a limited ability to correct prices by leaning against the demand of the two other types of investors.

We assume the following behavior of factor investors. In line with the vast literature on the flow-performance
relationship, we assume that these investors purchase additional amounts of the factor when it has performed well in the
 previous period. Past factor performance is given by the scalar $w'r_{t-1}$ and each stock is purchased in proportion $w_i$, so that net factor purchases are proportional to the vector $ww'r_{t-1}$. The price impact of these imbalances is then assumed to be proportional to trading so that

$$\left(\textrm{price impact due to flows into/out of factor investing}\right)_t = \alpha ww' r_{t-1}$$

\noindent This assumption is consistent with existing literature which documents that flow-induced trading has a
predictable component (corresponding to pure scaling up or down of the fund's portfolio i{}n response to changes in AUMs) and that this component has a measurable price impact (\cite{coval},\cite{GT}, \cite{lou}).

We end up with the following equations of motion for stock returns:

$${r}_{t} = {\mu} + {\epsilon}_{t} + \alpha \,w{w'} {r}_{t-1} $$

\noindent which implies the following dynamics of factor returns

$$F_t = \mu^{(F)} + a F_{t-1} + \epsilon^{(F)}_{t}$$

\noindent where we introduced $a=\alpha {w'} . {w}>0$ which is a scalar. The law of motion of factor returns implies
that there is autocorrelation of factor returns due to $a$ (performance-flow induced trading). $\mu^{(F)}=w'\mu$ is the unconditional expected return of the factor.

Finally, we assume that the stochastic term, ${\epsilon}_{t}$, to be strongly mean-reverting. We assume its dynamics is
governed by an MA1 process

$${\epsilon}_{t} = {e}_t-\rho {e}_{t-1}$$

\noindent where $\rho$ is a scalar and ${e}_{t} \sim {N}(0,\Sigma)$, where $\Sigma$ is the covariance matrix. Strong mean-reversion at fast time scales is an empirical
fact for stocks. It is thus a short-memory process, while returns themselves, through positive feedback trading, have a
longer memory.

\subsection{Solving the Model \& Results}

Define $A=\alpha ww'$. Stationarity implies that $a<1$, which is an assumption we will be making. Then, returns can be shown to be described by the following equation:

$$r_t = \frac{1}{1-A}\mu + e_t + (A-\rho I)e_{t-1}+ (a-\rho)\sum_{k \geq 2}{a^{k-2}Ae_{t-k}}$$

\noindent where the terms in $A-\rho I$ and $(a-\rho)$ represent the opposing forces of flow-induced trading ($a$) and noise trader demand ($\rho$). 

The autocovariance matrix of returns (of order $k$) is given by:

\begin{align*}
\Omega_1 &= (1-\rho(a-\rho)) A \Sigma - \rho \Sigma + (a-\rho) A \Sigma A \left( 1 + \frac{(a-\rho) \,a}{1-a^2}\right) \\
\Omega_k &= (a-\rho)(1-\rho a) a^{k-2} A \Sigma + (a-\rho) a^{k-1} A\Sigma A \left(1+\frac{(a-\rho)\,a}{1-a^2} \right) \text{ if }k\geq 2
\end{align*}

\bigskip

Using these formulas, it is straightforward to show our first result:

\begin{proposition}
\emph{Factor momentum PNL} \\
Let $V = w' \Sigma w$ denote the expected factor variance. Then, the expected PNL of factor momentum is given by:
$$E(F_{t-k}F_t)=w'\Omega_k w + \left(\mu^{(F)}\right)^2 = V \left(\frac{(a-\rho)(1-\rho a)}{1-a^2}\right)a^{k-1} + \left(\mu^{(F)}\right)^2$$

Thus, after controlling for mechanical exposure to factor investing ($\left(\mu^{(F)}\right)^2$), there is factor momentum \underline{at all lags} if and only if $a>\rho$
\end{proposition}

In the second term of the above expressions, we rediscover that the expected PNL of factor momentum is positive even if there is no flow-induced trading ($a=0$). This is a result we have seen previously: There is a mechanical relation between factor momentum and outright factor exposure.

But the interesting part of the above proposition is in the first term, which is the performance of factor momentum
\emph{in excess of mechanical exposure to factor investing}. This first term has the same sign at all horizons: If
excess return of factor momentum is positive for $k=1$, it will be positive at all horizons. This is
consistent with what we find in the data (Table \ref{fig:sr_stock_factor_mom_ts}).

Let us now turn to stock momentum. Stock momentum is defined thourgh the trace of the autocovariance matrices

\begin{proposition}
\emph{Stock momentum PNL} \\
Write $V=w'\Sigma w$. The expected PNL of stock momentum at all horizons $k$ is given by:
$$E(r_{t-1}r_t)=\text{tr}\left(\Omega_1\right) + \mu'.\mu =  \alpha V \left(1+\frac{(a-\rho)^2}{1-a^2}\right) - \rho \text{tr} \Sigma + \mu'.\mu$$
$$E(r_{t-k}r_t)=\text{tr}\left(\Omega_k\right) + \mu'.\mu = \alpha V \left(\frac{(a-\rho)(1-\rho a)}{1-a^2}\right)
a^{k-2} + \mu'.\mu \textrm{\,\,if\,\,}k\geq 2$$
Thus:
\begin{itemize}
	\item There is stock momentum at all horizons $k\geq2$ as long as $a>\rho$
	\item When $-\rho \text{tr} \Sigma$ starts dominating, there is mean-reversion for $k=1$
\end{itemize}
\end{proposition}

This second proposition demonstrates that, whenever there is factor momentum (i.e. $a>\rho$), there is also stock momentum for all $k \geq 2$. In this sense factor momentum and stock momentum are indistinguishable: They occur for the same parameter values as soon as one excludes the most recent lag.

Another inference is that when the stock universe becomes large (keeping $\rho$ constant), $-\rho \text{tr} \Sigma$
becomes dominating and stock momentum becomes mean reversion for $k=1$.

One last outcome of this analysis is that, stock momentum PNL ($\sim \text{tr}(\Omega_k)$) is a natural correlate of
factor momentum if the values of $a$ or $\rho$ change. The expression below

$$\alpha \,V\,\frac{(a-\rho)(1-\rho a)}{1-a^2}a^{k-2}$$

\noindent is present both in stock momentum (for $k \geq 2$) and in factor momentum. Variations in the aforementioned
parameters tend to make the two strategies comove. This is because, in our model, stock momentum is subsumed in factor
momentum (for $k \geq 2$). An interesting extension of this model would be to introduce ``idiosyncratic'' stock
momentum, for instance through assuming that $\epsilon_t$ follows an MA2 process with a positive loading on $e_{t-2}$.

\section{Conclusion}
\label{conclu}

In this paper we have established that, for most implementations, factor momentum is fully accounted for by stock momentum and factor exposure. An important exception occurs at one month lag (and various holding periods). Including the last month of stock returns in the definition of stock momentum degrades its performance because stocks exhibit mean reversion at monthly time scale. This is not true for factor momentum, as factors returns are persistent at all time scales that we have studied. Consequently, ``true'' factor momentum only exists through the effect of the last month of returns. If we exclude it, factor momentum is pure stock momentum.

It is possible to reconcile these empirical observations with a simple model, which we provide and solve. This model predicts that both strategies (factor and stock momentum) are correlated between each other to a constant level, independently of the lag and the holding period, consistent with evidence that we find empirically.

\newpage

\bibliography{zoo_bib.bbl}

\begin{thebibliography}{53}
\newcommand{\enquote}[1]{``#1''}
\expandafter\ifx\csname natexlab\endcsname\relax\def\natexlab#1{#1}\fi

\bibitem[\protect\citeauthoryear{Abarbanell and Bushee}{Abarbanell and
  Bushee}{1998}]{abarbanell1998}
\textsc{Abarbanell, J.~S. and B.~J. Bushee} (1998): \enquote{Abnormal Returns
  to a Fundamental Analysis Strategy,} \emph{The Accounting Review}, 73,
  19--45.

\bibitem[\protect\citeauthoryear{Ali, Hwang, and Trombley}{Ali
  et~al.}{2003}]{ali2003}
\textsc{Ali, A., L.-S. Hwang, and M.~A. Trombley} (2003): \enquote{Arbitrage
  risk and the book-to-market anomaly,} \emph{Journal of Financial Economics},
  69, 355--373.

\bibitem[\protect\citeauthoryear{Amihud}{Amihud}{2002}]{amihud2002}
\textsc{Amihud, Y.} (2002): \enquote{Illiquidity and stock returns:
  cross-section and time-series effects,} \emph{Journal of Financial Markets},
  5, 31--56.

\bibitem[\protect\citeauthoryear{Anderson and Garcia-Feijoo}{Anderson and
  Garcia-Feijoo}{2006}]{anderson2006}
\textsc{Anderson, C.~W. and L.~Garcia-Feijoo} (2006): \enquote{Empirical
  Evidence on Capital Investment, Growth Options, and Security Returns,}
  \emph{The Journal of Finance}, 61, 171--194.

\bibitem[\protect\citeauthoryear{Ang, Hodrick, Xing, and Zhang}{Ang
  et~al.}{2006}]{ang2006}
\textsc{Ang, A., R.~J. Hodrick, Y.~Xing, and X.~Zhang} (2006): \enquote{The
  Cross-Section of Volatility and Expected Returns,} \emph{The Journal of
  Finance}, 61, 259--299.

\bibitem[\protect\citeauthoryear{Asness, Porter, and Stevens}{Asness
  et~al.}{2000}]{asness2000}
\textsc{Asness, C.~S., R.~B. Porter, and R.~L. Stevens} (2000):
  \enquote{Predicting Stock Returns Using Industry-Relative Firm
  Characteristics,} .

\bibitem[\protect\citeauthoryear{Barbee, Mukherji, and Raines}{Barbee
  et~al.}{1996}]{barbee1996}
\textsc{Barbee, W.~C., S.~Mukherji, and G.~A. Raines} (1996): \enquote{Do
  Sales--Price and Debt--Equity Explain Stock Returns Better than Book--Market
  and Firm Size?} \emph{Financial Analysts Journal}, 52, 56--60.

\bibitem[\protect\citeauthoryear{Basu}{Basu}{1977}]{basu1977}
\textsc{Basu, S.} (1977): \enquote{Investment Performance of Common Stocks in
  Relation to Their Price-Earnings Ratios: A Test of the Efficient Market
  Hypothesis,} \emph{The Journal of Finance}, 32, 663--682.

\bibitem[\protect\citeauthoryear{Bhandari}{Bhandari}{1988}]{bhandari1988}
\textsc{Bhandari, L.~C.} (1988): \enquote{Debt/Equity Ratio and Expected Common
  Stock Returns: Empirical Evidence,} \emph{The Journal of Finance}, 43,
  507--528.

\bibitem[\protect\citeauthoryear{Campbell and Shiller}{Campbell and
  Shiller}{1987}]{CS}
\textsc{Campbell, J. and R.~Shiller} (1987): \enquote{Stock Prices, Earnings
  and Expected Dividends,} \emph{Journal of Finance}.

\bibitem[\protect\citeauthoryear{Chordia, Subrahmanyam, and Anshuman}{Chordia
  et~al.}{2001}]{chordia2001}
\textsc{Chordia, T., A.~Subrahmanyam, and V.~R. Anshuman} (2001):
  \enquote{Trading activity and expected stock returns,} \emph{Journal of
  Financial Economics}, 59, 3--32.

\bibitem[\protect\citeauthoryear{Cohen, Polk, and Vuolteenhao}{Cohen
  et~al.}{2003}]{cohen}
\textsc{Cohen, Polk, and Vuolteenhao} (2003): \enquote{The Value Spread,}
  \emph{Journal of Finance}.

\bibitem[\protect\citeauthoryear{Coval and Stafford}{Coval and
  Stafford}{2007}]{coval}
\textsc{Coval, J. and E.~Stafford} (2007): \enquote{Asset Fire Sales (and
  Purchases) on Equity Markets,} \emph{Journal of Financial Economics}.

\bibitem[\protect\citeauthoryear{Daniel and Titman}{Daniel and
  Titman}{2006}]{daniel2006}
\textsc{Daniel, K. and S.~Titman} (2006): \enquote{Market Reactions to Tangible
  and Intangible Information,} \emph{The Journal of Finance}, 61, 1605--1643.

\bibitem[\protect\citeauthoryear{De~Bondt and Thaler}{De~Bondt and
  Thaler}{1985}]{debondt1985}
\textsc{De~Bondt, W. F.~M. and R.~Thaler} (1985): \enquote{Does the Stock
  Market Overreact?} \emph{The Journal of Finance}, 40, 793--805.

\bibitem[\protect\citeauthoryear{Desai, Rajgopal, and Venkatachalam}{Desai
  et~al.}{2004}]{desai2004}
\textsc{Desai, H., S.~Rajgopal, and M.~Venkatachalam} (2004):
  \enquote{Values-Glamour and Accruals Mispricing: One Anomaly or Two?}
  \emph{The Accounting Review}, 79, 355--385.

\bibitem[\protect\citeauthoryear{Ehsani and Linnainmaa}{Ehsani and
  Linnainmaa}{2019}]{M2}
\textsc{Ehsani and Linnainmaa} (2019): \enquote{Factor momentum and the
  momentum factor,} Tech. rep.

\bibitem[\protect\citeauthoryear{Fama and French}{Fama and
  French}{1993}]{fama1993}
\textsc{Fama, E.~F. and K.~R. French} (1993): \enquote{Common risk factors in
  the returns on stocks and bonds,} \emph{Journal of Financial Economics}, 33,
  3--56.

\bibitem[\protect\citeauthoryear{Francis, LaFond, Olsson, and Schipper}{Francis
  et~al.}{2004}]{francis2004}
\textsc{Francis, J., R.~LaFond, P.~M. Olsson, and K.~Schipper} (2004):
  \enquote{Costs of Equity and Earnings Attributes,} \emph{The Accounting
  Review}, 79, 967--1010.

\bibitem[\protect\citeauthoryear{Giglio, Feng, and Xiu}{Giglio
  et~al.}{2020}]{GFX}
\textsc{Giglio, Feng, and Xiu} (2020): \enquote{Taming the Factor Zoo: a Test
  of New Factors,} \emph{Journal of Finance}.

\bibitem[\protect\citeauthoryear{Greenwood and Thesmar}{Greenwood and
  Thesmar}{2011}]{GT}
\textsc{Greenwood, R. and D.~Thesmar} (2011): \enquote{Stock Price Fragility,}
  \emph{Journal of Financial Economics}.

\bibitem[\protect\citeauthoryear{Haddad, Kozak, and Santosh}{Haddad
  et~al.}{2020}]{valentin}
\textsc{Haddad, Kozak, and Santosh} (2020): \enquote{Factor Timing,}
  \emph{Review of Financial Studies}.

\bibitem[\protect\citeauthoryear{Hanson and Sunderam}{Hanson and
  Sunderam}{2014}]{hanson}
\textsc{Hanson, S. and A.~Sunderam} (2014): \enquote{The Growth and Limits of
  Arbitrage: Evidence from Short Interest,} \emph{Review of Financial Studies}.

\bibitem[\protect\citeauthoryear{Harvey and Zhu}{Harvey and Zhu}{2016}]{...and}
\textsc{Harvey, L. and Zhu} (2016): \enquote{... and the cross-section of
  Expected Returns,} \emph{Review of Financial Studies}.

\bibitem[\protect\citeauthoryear{Haugen and Baker}{Haugen and
  Baker}{1996}]{haugen1996}
\textsc{Haugen, R.~A. and N.~L. Baker} (1996): \enquote{Commonality in the
  determinants of expected stock returns,} \emph{Journal of Financial
  Economics}, 41, 401--439.

\bibitem[\protect\citeauthoryear{Heston and Sadka}{Heston and
  Sadka}{2008}]{heston2008}
\textsc{Heston, S.~L. and R.~Sadka} (2008): \enquote{Seasonality in the
  cross-section of stock returns,} \emph{Journal of Financial Economics}, 87,
  418--445.

\bibitem[\protect\citeauthoryear{Holthausen and Larcker}{Holthausen and
  Larcker}{1992}]{holthausen1992}
\textsc{Holthausen, R.~W. and D.~F. Larcker} (1992): \enquote{The prediction of
  stock returns using financial statement information,} \emph{Journal of
  Accounting and Economics}, 15, 373--411.

\bibitem[\protect\citeauthoryear{Hou and Moskowitz}{Hou and
  Moskowitz}{2005}]{hou2005}
\textsc{Hou, K. and T.~J. Moskowitz} (2005): \enquote{Market Frictions, Price
  Delay and the Cross-Section of Expected Returns,} \emph{The Review of
  Financial Studies}, 18, 981--1020.

\bibitem[\protect\citeauthoryear{Hou and Robinson}{Hou and
  Robinson}{2006}]{hou2006}
\textsc{Hou, K. and D.~T. Robinson} (2006): \enquote{Industry Concentration and
  Average Stock Returns,} \emph{The Journal of Finance}, 61, 1927--1956.

\bibitem[\protect\citeauthoryear{Hou and Zhang}{Hou and Zhang}{2019}]{HXZ}
\textsc{Hou, X. and .~Zhang} (2019): \enquote{Replicating anomalies,}
  \emph{Review of Financial Studies}.

\bibitem[\protect\citeauthoryear{Jegadeesh and Livnat}{Jegadeesh and
  Livnat}{2006}]{jegadeesh2006}
\textsc{Jegadeesh, N. and J.~Livnat} (2006): \enquote{Revenue surprises and
  stock returns,} \emph{Journal of Accounting and Economics}, 41, 147--171.

\bibitem[\protect\citeauthoryear{Jegadeesh and Titman}{Jegadeesh and
  Titman}{1993}]{jegadeesh1993}
\textsc{Jegadeesh, N. and S.~Titman} (1993): \enquote{Returns to Buying Winners
  and Selling Losers: Implications for Stock Market Efficiency,} \emph{The
  Journal of Finance}, 48, 65--91.

\bibitem[\protect\citeauthoryear{Jiang, Lee, and Zhang}{Jiang
  et~al.}{2005}]{jiang2005}
\textsc{Jiang, G., C.~M.~C. Lee, and Y.~Zhang} (2005): \enquote{Information
  Uncertainty and Expected Returns,} \emph{Review of Accounting Studies}, 10,
  185--221.

\bibitem[\protect\citeauthoryear{Lakonishok, Shleifer, and Vishny}{Lakonishok
  et~al.}{1994}]{lakonishok1994}
\textsc{Lakonishok, J., A.~Shleifer, and R.~W. Vishny} (1994):
  \enquote{Contrarian investment, extrapolation, and risk,} \emph{The Journal
  of Finance}, 49, 1541--1578.

\bibitem[\protect\citeauthoryear{Lev and Nissim}{Lev and
  Nissim}{2004}]{lev2004}
\textsc{Lev, B. and D.~Nissim} (2004): \enquote{Taxable Income, Future
  Earnings, and Equity Values,} \emph{The Accounting Review}, 79, 1039--1074.

\bibitem[\protect\citeauthoryear{Liu}{Liu}{2006}]{liu2006}
\textsc{Liu, W.} (2006): \enquote{A liquidity-augmented capital asset pricing
  model,} \emph{Journal of Financial Economics}, 82, 631--671.

\bibitem[\protect\citeauthoryear{Lo and MacKinlay}{Lo and MacKinlay}{1990}]{lo}
\textsc{Lo, A. and C.~MacKinlay} (1990): \enquote{When are contrarian profits
  due to stock market overreaction?} \emph{The Review of Financial Studies}, 3,
  175--205.

\bibitem[\protect\citeauthoryear{Lou}{Lou}{2012}]{lou}
\textsc{Lou, D.} (2012): \enquote{Flow-Based Explanation for Return
  Predictability,} \emph{Review of Financial Studies}.

\bibitem[\protect\citeauthoryear{Lyandres, Sun, and Zhang}{Lyandres
  et~al.}{2008}]{lyandres2008}
\textsc{Lyandres, E., L.~Sun, and L.~Zhang} (2008): \enquote{The New Issues
  Puzzle: Testing the Investment-Based Explanation,} \emph{The Review of
  Financial Studies}, 21, 2825--2855.

\bibitem[\protect\citeauthoryear{McLean and Pontiff}{McLean and
  Pontiff}{2016}]{pontiff}
\textsc{McLean and Pontiff} (2016): \enquote{Does academic research destroy
  stock return predictability?} \emph{Journal of Finance}.

\bibitem[\protect\citeauthoryear{Moskowitz and Grinblatt}{Moskowitz and
  Grinblatt}{1999}]{moskowitz1999}
\textsc{Moskowitz, T.~J. and M.~Grinblatt} (1999): \enquote{Do Industries
  Explain Momentum?} \emph{The Journal of Finance}, 54, 1249--1290.

\bibitem[\protect\citeauthoryear{Novy-Marx}{Novy-Marx}{2012}]{novymom}
\textsc{Novy-Marx, R.} (2012): \enquote{Is Momentum Really Momentum?}
  \emph{Journal of Financial Economics}, 103, 429--453.

\bibitem[\protect\citeauthoryear{Ou and Penman}{Ou and Penman}{1989}]{ou1989}
\textsc{Ou, J.~A. and S.~H. Penman} (1989): \enquote{Financial Statement
  Analysis and the Prediction of Stock Returns,} \emph{Journal of Accounting
  and Economics}, 11, 295--329.

\bibitem[\protect\citeauthoryear{Pastor and Stambaugh}{Pastor and
  Stambaugh}{2003}]{pastor2003}
\textsc{Pastor, L. and R.~F. Stambaugh} (2003): \enquote{Liquidity Risk and
  Expected Stock Returns,} \emph{Journal of Political Economy}, 111, 642--685.

\bibitem[\protect\citeauthoryear{Piotroski}{Piotroski}{2000}]{piotroski2000}
\textsc{Piotroski, J.~D.} (2000): \enquote{Value Investing: The Use of
  Historical Financial Statement Information to Separate Winners from Losers,}
  \emph{Journal of Accounting Research}, 38, 1--41.

\bibitem[\protect\citeauthoryear{Pontiff and Woodgate}{Pontiff and
  Woodgate}{2008}]{pontiff2008}
\textsc{Pontiff, J. and A.~Woodgate} (2008): \enquote{Share Issuance and
  Cross-sectional Returns,} \emph{The Journal of Finance}, 63, 921--945.

\bibitem[\protect\citeauthoryear{Rendleman, Jones, and Latan{\'e}}{Rendleman
  et~al.}{1982}]{rendleman1982}
\textsc{Rendleman, R.~J., C.~P. Jones, and H.~A. Latan{\'e}} (1982):
  \enquote{Empirical anomalies based on unexpected earnings and the importance
  of risk adjustments,} \emph{Journal of Financial Economics}, 10, 269--287.

\bibitem[\protect\citeauthoryear{Richardson, Sloan, Soliman, and
  Tuna}{Richardson et~al.}{2005}]{richardson2005}
\textsc{Richardson, S.~A., R.~G. Sloan, M.~T. Soliman, and I.~Tuna} (2005):
  \enquote{Accrual reliability, earnings persistence and stock prices,}
  \emph{Journal of Accounting and Economics}, 39, 437--485.

\bibitem[\protect\citeauthoryear{Sloan}{Sloan}{1996}]{Sloan1996}
\textsc{Sloan, R.~G.} (1996): \enquote{Do Stock Prices Fully Reflect
  Information in Accruals and Cash Flows about Future Earnings?} \emph{The
  Accounting Review}, 71, 289--315.

\bibitem[\protect\citeauthoryear{Soliman}{Soliman}{2008}]{soliman2008}
\textsc{Soliman, M.~T.} (2008): \enquote{The Use of DuPont Analysis by Market
  Participants,} \emph{The Accounting Review}, 83, 823--853.

\bibitem[\protect\citeauthoryear{Thomas and Zhang}{Thomas and
  Zhang}{2002}]{thomas2002}
\textsc{Thomas, J.~K. and H.~Zhang} (2002): \enquote{Inventory Changes and
  Future Returns,} \emph{Review of Accounting Studies}, 7, 163--187.

\bibitem[\protect\citeauthoryear{Titman, Wei, and Xie}{Titman
  et~al.}{2004}]{titman2004}
\textsc{Titman, S., J.~K.~C. Wei, and F.~Xie} (2004): \enquote{Capital
  Investments and Stock Returns,} \emph{Journal of Financial and Quantitative
  Analysis}, 39, 677--700.

\bibitem[\protect\citeauthoryear{Xing}{Xing}{2008}]{xing2008}
\textsc{Xing, Y.} (2008): \enquote{Interpreting the Value Effect Through the
  Q-Theory: An Empirical Investigation,} \emph{The Review of Financial
  Studies}, 21, 1767--1795.

\end{thebibliography}






@article{coval,
	Author = {Joshua Coval and Erik Stafford},
	Date-Added = {2020-08-20 23:13:58 +0200},
	Date-Modified = {2020-08-20 23:14:30 +0200},
	Journal = {Journal of Financial Economics},
	Title = {Asset Fire Sales (and Purchases) on Equity Markets},
	Year = {2007}}


@article{GT,
	Author = {Robin Greenwood and David Thesmar},
	Date-Added = {2020-08-20 23:13:58 +0200},
	Date-Modified = {2020-08-20 23:14:30 +0200},
	Journal = {Journal of Financial Economics},
	Title = {Stock Price Fragility},
	Year = {2011}}

@article{lou,
	Author = {Dong Lou},
	Date-Added = {2020-08-20 23:13:13 +0200},
	Date-Modified = {2020-08-20 23:16:40 +0200},
	Journal = {Review of Financial Studies},
	Title = {Flow-Based Explanation for Return Predictability},
	Year = {2012}}


@article{DSSW,
	Author = {De Long and Shleifer and Summers and Waldmann},
	Date-Added = {2020-07-01 17:39:26 -0400},
	Date-Modified = {2020-07-01 17:40:43 -0400},
	Journal = {Journal of Political Economy},
	Title = {Noise Trader Risk and Financial Markets},
	Year = {1990}}

@article{DSSW2,
	Author = {Bradford De Long and Andrei Shleifer and Lawrence Summers and Robert Waldmann},
	Date-Added = {2020-06-28 18:13:27 -0400},
	Date-Modified = {2020-06-28 18:14:38 -0400},
	Journal = {Journal of Finance},
	Title = {Positive Feedback Investment Strategies and Destabilizing Rational Speculation},
	Year = {1990}}

@article{hongstein,
	Author = {Harrison Hong and Jeremy Stein},
	Date-Added = {2020-06-28 17:55:09 -0400},
	Date-Modified = {2020-06-28 17:55:32 -0400},
	Journal = {Journal of Finance},
	Title = {A Unified Theory of Underreaction, Momentum Trading and Overreaction in Asset Markets},
	Year = {1999}}

@article{hanson,
	Author = {Sam Hanson and Adi Sunderam},
	Date-Added = {2020-06-26 19:04:01 -0400},
	Date-Modified = {2020-06-26 19:06:07 -0400},
	Journal = {Review of Financial Studies},
	Title = {The Growth and Limits of Arbitrage: Evidence from Short Interest},
	Year = {2014}}

@article{CS,
	Author = {John Campbell and Robert Shiller},
	Date-Added = {2020-06-26 18:57:04 -0400},
	Date-Modified = {2020-06-26 18:58:08 -0400},
	Journal = {Journal of Finance},
	Title = {Stock Prices, Earnings and Expected Dividends},
	Year = {1987}}

@article{novymom,
	Author = {Robert Novy-Marx},
	Date-Added = {2020-06-03 16:06:06 -0400},
	Date-Modified = {2020-06-03 16:07:12 -0400},
	Journal = {Journal of Financial Economics},
	Pages = {429-453},
	Title = {Is Momentum Really Momentum?},
	Volume = {103},
	Year = {2012}}

@article{lo,
	Author = {Andrew Lo and Craig MacKinlay},
	Journal = {The Review of Financial Studies},
	Pages = {175-205},
	Title = {When are contrarian profits due to stock market overreaction?},
	Volume = 3,
	Year = 1990}

@article{ross,
	Author = {Stephen Ross},
	Journal = {Journal of Economic Theory},
	Pages = {341-360},
	Title = {The arbitrage theory of capital asset pricing},
	Volume = 13,
	Year = 1976}

@article{tuzel2010,
	Author = {Tuzel, Selale},
	Journal = {The Review of Financial Studies},
	Keywords = {factor-investing real-estate},
	Number = 6,
	Pages = {2268-2302},
	Title = {Corporate Real Estate Holdings and the Cross-Section of Stock Returns},
	Volume = 23,
	Year = 2010}

@unpublished{bandyopadhyay2010,
	Author = {Bandyopadhyay, Sati P. and Huang, Alan G. and Wirjanto, Tony S.},
	Keywords = {accrual-volatility earnings factor-investing},
	Title = {The Accrual Volatility Anomaly},
	Year = 2010}

@article{huang2009,
	Author = {Huang, Alan Guoming},
	Journal = {Journal of Empirical Finance},
	Month = jan,
	Number = 1,
	Pages = {409-429},
	Title = {The cross section of cashflow volatility and expected stock returns},
	Volume = 16,
	Year = 2009}

@article{hong2009,
	Author = {Hong, Harrison and Kacperczyk, Marcin},
	Journal = {Journal of Financial Economics},
	Month = {April},
	Number = 1,
	Pages = {15-36},
	Title = {The price of sin: The effects of social norms on markets},
	Volume = 93,
	Year = 2009}

@preprint{chandrashekar2009,
	Author = {Chandrashekar, Satyajit and Rao, Ramesh K. S.},
	Keywords = {cash-productivity factor-investing},
	Title = {The Productivity of Cash and the Cross-Section of Expected Stock Returns},
	Year = 2009}

@article{xing2008,
	Author = {Xing, Yuhang},
	Journal = {The Review of Financial Studies},
	Keywords = {factor-investing investment investment-growth q-theory},
	Month = {July},
	Number = 4,
	Pages = {1767-1795},
	Title = {Interpreting the Value Effect Through the Q-Theory: An Empirical Investigation},
	Volume = 21,
	Year = 2008}

@article{soliman2008,
	Author = {Soliman, Mark T.},
	Journal = {The Accounting Review},
	Month = may,
	Number = 3,
	Pages = {823-853},
	Title = {The Use of DuPont Analysis by Market Participants},
	Volume = 83,
	Year = 2008}

@article{lyandres2008,
	Author = {Lyandres, Evgeny and Sun, Le and Zhang, Lu},
	Journal = {The Review of Financial Studies},
	Month = nov,
	Number = 6,
	Pages = {2825--2855},
	Title = {The New Issues Puzzle: Testing the Investment-Based Explanation},
	Volume = 21,
	Year = 2008}

@article{heston2008,
	Author = {Heston, Steven L. and Sadka, Ronnie},
	Journal = {Journal of Financial Economics},
	Month = feb,
	Number = 2,
	Pages = {418-445},
	Title = {Seasonality in the cross-section of stock returns},
	Volume = 87,
	Year = 2008}

@preprint{brandt2008,
	Author = {Brandt, Michael W. and Kishore, Runeet and Santa-Clara, Pedro and Venkatachalam, Mohan},
	Title = {Earnings Announcements are Full of Surprises},
	Year = 2008}

@preprint{lerman2008,
	Author = {Lerman, Alina and Livnat, Joshua and Mendenhall, Richard R.},
	Title = {The High-Volume Return Premium and Post-Earnings Announcement Drift},
	Year = 2008}

@article{pontiff2008,
	Author = {Pontiff, Jeffrey and Woodgate, Artemiza},
	Journal = {The Journal of Finance},
	Month = {April},
	Number = 2,
	Pages = {921-945},
	Title = {Share Issuance and Cross-sectional Returns},
	Volume = 63,
	Year = 2008}

@article{lamont2001,
	Author = {Lamont, Owen and Polk, Christopher and Saa-Requejo, Jesus},
	Journal = {The Review of Financial Studies},
	Number = 2,
	Pages = {529-554},
	Title = {Financial Constraints and Stock Returns},
	Volume = 14,
	Year = 2001}

@article{chan2001,
	Added-At = {2019-06-06T17:30:43.000+0200},
	Author = {Chan, Louis K. C. and Lakonishok, Josef and Sougiannis, Theodore},
	Journal = {The Journal of Finance},
	Month = dec,
	Number = 6,
	Pages = {2431-2456},
	Title = {The Stock Market Valuation of Research and Development Expenditures},
	Volume = 56,
	Year = 2001}

@article{penman2007,
	Author = {Penman, Stephen H. and Richardson, Scott A. and Tuna, Irem},
	Journal = {Journal of Accounting Research},
	Month = may,
	Number = 2,
	Pages = {427-467},
	Title = {The Book-to-Price Effect in Stock Returns: Accounting for Leverage},
	Volume = 45,
	Year = 2007}

@article{boudoukh2007,
	Author = {Boudoukh, Jacob and Michaely, Roni and Richardson, Matthew and Roberts, Michael R.},
	Journal = {The Journal of Finance},
	Month = apr,
	Number = 2,
	Pages = {877-915},
	Title = {On the Importance of Measuring Payout Yield: Implications for Empirical Asset Pricing},
	Volume = 62,
	Year = 2007}

@article{almeida2007,
	Author = {Almeida, Heitor and Campello, Murillo},
	Journal = {The Review of Financial Studies},
	Month = sep,
	Number = 5,
	Pages = {1429-1460},
	Title = {Financial Constraints, Asset Tangibility, and Corporate Investment},
	Volume = 20,
	Year = 2007}

@article{whited2006,
	Author = {Whited, Toni M. and Wu, Guojun},
	Journal = {The Review of Financial Studies},
	Month = jan,
	Number = 2,
	Pages = {531-559},
	Title = {Financial Constraints Risk},
	Volume = 19,
	Year = 2006}

@article{hou2006,
	Author = {Hou, Kewei and Robinson, David T.},
	Journal = {The Journal of Finance},
	Month = aug,
	Number = 4,
	Pages = {1927-1956},
	Title = {Industry Concentration and Average Stock Returns},
	Volume = 61,
	Year = 2006}

@article{jegadeesh2006,
	Author = {Jegadeesh, Narasimhan and Livnat, Joshua},
	Journal = {Journal of Accounting and Economics},
	Pages = {147-171},
	Title = {Revenue surprises and stock returns},
	Volume = 41,
	Year = 2006}

@article{bradshaw2006,
	Author = {Bradshaw, Mark T. and Richardson, Scott A. and Sloan, Richard G.},
	Journal = {Journal of Accounting and Economics},
	Number = {53-85},
	Title = {The relation between corporate financing activities, analysts' forecasts and stock returns},
	Volume = 42,
	Year = 2006}

@article{daniel2006,
	Author = {Daniel, Kent and Titman, Sheridan},
	Journal = {The Journal of Finance},
	Month = aug,
	Number = 4,
	Pages = {1605-1643},
	Title = {Market Reactions to Tangible and Intangible Information},
	Volume = 61,
	Year = 2006}

@article{liu2006,
	Author = {Liu, Weimin},
	Journal = {Journal of Financial Economics},
	Month = {June},
	Pages = {631-671},
	Title = {A liquidity-augmented capital asset pricing model},
	Volume = 82,
	Year = 2006}

@article{anderson2006,
	Author = {Anderson, Christopher W. and Garcia-Feijoo, Luis},
	Journal = {The Journal of Finance},
	Month = feb,
	Number = 1,
	Pages = {171-194},
	Title = {Empirical Evidence on Capital Investment, Growth Options, and Security Returns},
	Volume = 61,
	Year = 2006}

@article{mohanram2005,
	Author = {Mohanram, S. Partha},
	Journal = {Review of Accounting Studies},
	Pages = {133-170},
	Title = {Separating Winners from Losers among Low Book-to-Market Stocks using Financial Statement Analysis},
	Volume = 10,
	Year = 2005}

@article{ang2006,
	Author = {Ang, Andrew and Hodrick, Robert J. and Xing, Yuhang and Zhang, Xiaoyan},
	Journal = {The Journal of Finance},
	Month = feb,
	Number = 1,
	Pages = {259-299},
	Title = {The Cross-Section of Volatility and Expected Returns},
	Volume = 61,
	Year = 2006}

@article{richardson2005,
	Author = {Richardson, Scott A. and Sloan, Richard G. and Soliman, Mark T. and Tuna, Irem},
	Journal = {Journal of Accounting and Economics},
	Pages = {437-485},
	Title = {Accrual reliability, earnings persistence and stock prices},
	Volume = 39,
	Year = 2005}

@article{jiang2005,
	Author = {Jiang, Guohua and Lee, Charles M. C. and Zhang, Yi},
	Journal = {Review of Accounting Studies},
	Pages = {185-221},
	Title = {Information Uncertainty and Expected Returns},
	Volume = 10,
	Year = 2005}

@article{hou2005,
	Author = {Hou, Kewei and Moskowitz, Tobias J.},
	Journal = {The Review of Financial Studies},
	Number = 3,
	Pages = {981-1020},
	Title = {Market Frictions, Price Delay and the Cross-Section of Expected Returns},
	Volume = 18,
	Year = 2005}

@article{lev2004,
	Author = {Lev, Baruch and Nissim, Doron},
	Journal = {The Accounting Review},
	Number = 4,
	Pages = {1039-1074},
	Title = {Taxable Income, Future Earnings, and Equity Values},
	Volume = 79,
	Year = 2004}

@article{hirshleifer2004,
	Author = {Hirshleifer, David and Hou, Kewei and Hong Teoh, Siew and Zhang, Yinglei},
	Journal = {Journal of Accounting and Economics},
	Pages = {297-331},
	Title = {Do investors overvalue firms with bloated balance sheets?},
	Volume = 38,
	Year = 2004}

@article{francis2004,
	Author = {Francis, Jennifer and LaFond, Ryan and Olsson, Per M. and Schipper, Katherine},
	Journal = {The Accounting Review},
	Month = oct,
	Number = 4,
	Pages = {967-1010},
	Title = {Costs of Equity and Earnings Attributes},
	Volume = 79,
	Year = 2004}

@article{titman2004,
	Author = {Titman, Sheridan and Wei, John K. C. and Xie, Feixue},
	Journal = {Journal of Financial and Quantitative Analysis},
	Month = dec,
	Number = 4,
	Pages = {677-700},
	Title = {Capital Investments and Stock Returns},
	Volume = 39,
	Year = 2004}

@article{eberhart2004,
	Author = {Eberhart, Allan C. and Maxwell, William F. and Siddique, Akhtar R.},
	Journal = {The Journal of Finance},
	Month = apr,
	Number = 2,
	Pages = {623-650},
	Title = {An Examination of Long-Term Abnormal Stock Returns and Operating Performance Following R\&D Increases},
	Volume = 59,
	Year = 2004}

@article{desai2004,
	Author = {Desai, Hemang and Rajgopal, Shivaram and Venkatachalam, Mohan},
	Journal = {The Accounting Review},
	Number = 2,
	Pages = {355-385},
	Title = {Values-Glamour and Accruals Mispricing: One Anomaly or Two?},
	Volume = 79,
	Year = 2004}

@article{rajgopal2003,
	Author = {Rajgopal, Shivaram and Shevlin, Terry and Venkatakchalam, Mohan},
	Journal = {Review of Accounting Studies},
	Pages = {461-492},
	Title = {Does the Stock Market Fully Appreciate the Implications of Leading Indicators for Future Earnings? Evidence from Order Backlog},
	Volume = 8,
	Year = 2003}

@article{fairfield2003,
	Author = {Fairfield, P.M. and Whisenant, S. and Yohn, T.L.},
	Journal = {Review of Accounting Studies},
	Pages = {221-243},
	Title = {The Differential Persistence of Accruals and Cash Flows for Future Operating Income versus Future Profitability},
	Volume = 8,
	Year = 2003}

@article{ali2003,
	Author = {Ali, Ashiq and Hwang, Lee-Seok and Trombley, Mark A.},
	Journal = {Journal of Financial Economics},
	Month = aug,
	Number = 2,
	Pages = {355-373},
	Title = {Arbitrage risk and the book-to-market anomaly},
	Url = {http://www.sciencedirect.com/science/article/B6VBX-48N30X4-4/1/07df6f418e8b8ec6cdf83b5b97397a97},
	Volume = 69,
	Year = 2003,
	Bdsk-Url-1 = {http://www.sciencedirect.com/science/article/B6VBX-48N30X4-4/1/07df6f418e8b8ec6cdf83b5b97397a97}}

@article{pastor2003,
	Author = {Pastor, Lubos and Stambaugh, Robert F.},
	Journal = {Journal of Political Economy},
	Month = {June},
	Number = 3,
	Pages = {642-685},
	Title = {Liquidity Risk and Expected Stock Returns},
	Volume = 111,
	Year = 2003}

@article{amihud2002,
	Author = {Amihud, Yakov},
	Journal = {Journal of Financial Markets},
	Pages = {31-56},
	Title = {Illiquidity and stock returns: cross-section and time-series effects},
	Volume = 5,
	Year = 2002}

@article{thomas2002,
	Author = {Thomas, Jacob K. and Zhang, Huai},
	Journal = {Review of Accounting Studies},
	Pages = {163-187},
	Title = {Inventory Changes and Future Returns},
	Volume = 7,
	Year = 2002}

@article{chordia2001,
	Author = {Chordia, Tarun and Subrahmanyam, Avanidhar and Anshuman, V. Ravi},
	Journal = {Journal of Financial Economics},
	Month = jan,
	Number = 1,
	Pages = {3-32},
	Title = {Trading activity and expected stock returns},
	Url = {http://www.sciencedirect.com/science/article/B6VBX-41N5GP5-1/1/a58e4d9e4b1e40dfc3d6aac80a1f48a0},
	Volume = 59,
	Year = 2001,
	Bdsk-Url-1 = {http://www.sciencedirect.com/science/article/B6VBX-41N5GP5-1/1/a58e4d9e4b1e40dfc3d6aac80a1f48a0}}

@preprint{asness2000,
	Author = {Asness, Clifford S. and Porter, R. Burt and Stevens, Ross L.},
	Month = feb,
	Title = {Predicting Stock Returns Using Industry-Relative Firm Characteristics},
	Url = {https://papers.ssrn.com/sol3/papers.cfm?abstract_id=213872},
	Year = 2000,
	Bdsk-Url-1 = {https://papers.ssrn.com/sol3/papers.cfm?abstract_id=213872}}

@article{piotroski2000,
	Author = {Piotroski, Joseph D.},
	Doi = {10.2307/2672906},
	Journal = {Journal of Accounting Research},
	Pages = {1-41},
	Title = {Value Investing: The Use of Historical Financial Statement Information to Separate Winners from Losers},
	Url = {https://www.jstor.org/stable/2672906},
	Volume = 38,
	Year = 2000,
	Bdsk-Url-1 = {https://www.jstor.org/stable/2672906},
	Bdsk-Url-2 = {https://doi.org/10.2307/2672906}}

@article{moskowitz1999,
	Author = {Moskowitz, Tobias J. and Grinblatt, Mark},
	Doi = {https://doi.org/10.1111/0022-1082.00146},
	Journal = {The Journal of Finance},
	Month = {August},
	Number = 4,
	Pages = {1249-1290},
	Title = {Do Industries Explain Momentum?},
	Volume = 54,
	Year = 1999,
	Bdsk-Url-1 = {https://doi.org/10.1111/0022-1082.00146}}

@article{barth1999,
	Author = {Barth, Mary E. and Elliott, John A. and Finn, Mark W.},
	Journal = {Journal of Accounting Research},
	Number = 2,
	Pages = {387-413},
	Title = {Market Rewards Associated with Patterns of Increasing Earnings},
	Volume = 37,
	Year = 1999}

@article{dichev1998,
	Author = {Dichev, Ilia D.},
	Doi = {https://doi.org/10.1111/0022-1082.00046},
	Journal = {The Journal of Finance},
	Month = {June},
	Number = 3,
	Pages = {1131-1147},
	Title = {Is the Risk of Bankruptcy a Systemic Risk},
	Volume = 53,
	Year = 1998,
	Bdsk-Url-1 = {https://doi.org/10.1111/0022-1082.00046}}

@article{abarbanell1998,
	Author = {Abarbanell, Jeffery S. and Bushee, Brian J.},
	Journal = {The Accounting Review},
	Month = jan,
	Number = 1,
	Pages = {19-45},
	Title = {Abnormal Returns to a Fundamental Analysis Strategy},
	Volume = 73,
	Year = 1998}

@article{haugen1996,
	Author = {Haugen, Robert A. and Baker, Nardin L.},
	Doi = {https://doi.org/10.1016/0304-405X(95)00868-F},
	Journal = {Journal of Financial Economics},
	Month = jul,
	Number = 3,
	Pages = {401-439},
	Title = {Commonality in the determinants of expected stock returns},
	Url = {http://www.sciencedirect.com/science/article/B6VBX-3VVVRX5-8/1/07d7a025cca65a6f3286e34278eb79d0},
	Volume = 41,
	Year = 1996,
	Bdsk-Url-1 = {http://www.sciencedirect.com/science/article/B6VBX-3VVVRX5-8/1/07d7a025cca65a6f3286e34278eb79d0},
	Bdsk-Url-2 = {https://doi.org/10.1016/0304-405X(95)00868-F}}

@article{barbee1996,
	Author = {Barbee, William C. and Mukherji, S. and Raines, Gary A.},
	Doi = {10.2469/faj.v52.n2.1980},
	Journal = {Financial Analysts Journal},
	Number = 2,
	Pages = {56-60},
	Title = {Do Sales--Price and Debt--Equity Explain Stock Returns Better than Book--Market and Firm Size?},
	Volume = 52,
	Year = 1996,
	Bdsk-Url-1 = {https://doi.org/10.2469/faj.v52.n2.1980}}

@article{sloan1996,
	Author = {Sloan, Richard G.},
	Journal = {The Accounting Review},
	Month = jul,
	Number = 3,
	Pages = {289-315},
	Title = {Do Stock Prices Fully Reflect Information in Accruals and Cash Flows about Future Earnings?},
	Url = {https://www.jstor.org/stable/248290?seq=1#page_scan_tab_contents},
	Volume = 71,
	Year = 1996,
	Bdsk-Url-1 = {https://www.jstor.org/stable/248290?seq=1#page_scan_tab_contents}}

@article{roni1995,
	Author = {Michaely, Roni and Thaler, Richard H. and Womack, Kent},
	Doi = {10.1111/j.1540-6261.1995.tb04796.x},
	Journal = {The Journal of Finance},
	Month = {June},
	Number = 2,
	Pages = {573-608},
	Pagetotal = {39},
	Subtitle = {overreaction or drift?},
	Title = {Price reactions to dividend initiations and omissions: Overreaction or Drift?},
	Volume = 50,
	Year = 1995,
	Bdsk-Url-1 = {https://doi.org/10.1111/j.1540-6261.1995.tb04796.x}}

@article{loughran1995,
	Author = {Loughran, Tim and Ritter, Jay R.},
	Doi = {10.2307/2329238},
	Journal = {The Journal of Finance},
	Month = mar,
	Number = 1,
	Pages = {23-51},
	Title = {The New Issue Puzzle},
	Url = {https://www.jstor.org/stable/2329238?seq=1#page_scan_tab_contents},
	Volume = 50,
	Year = 1995,
	Bdsk-Url-1 = {https://www.jstor.org/stable/2329238?seq=1#page_scan_tab_contents},
	Bdsk-Url-2 = {https://doi.org/10.2307/2329238}}

@article{lakonishok1994,
	Author = {Lakonishok, Josef and Shleifer, Andrei and Vishny, Robert W.},
	Doi = {https://doi.org/10.1111/j.1540-6261.1994.tb04772.x},
	Journal = {The Journal of Finance},
	Month = dec,
	Number = 5,
	Pages = {1541-1578},
	Title = {Contrarian investment, extrapolation, and risk},
	Volume = 49,
	Year = 1994,
	Bdsk-Url-1 = {https://doi.org/10.1111/j.1540-6261.1994.tb04772.x}}

@article{jegadeesh1993,
	Author = {Jegadeesh, Narasimhan and Titman, Sheridan},
	Doi = {10.2307/2328882},
	Journal = {The Journal of Finance},
	Month = mar,
	Number = 1,
	Pages = {65-91},
	Title = {Returns to Buying Winners and Selling Losers: Implications for Stock Market Efficiency},
	Volume = 48,
	Year = 1993,
	Bdsk-Url-1 = {https://doi.org/10.2307/2328882}}

@article{fama1993,
	Author = {Fama, Eugene F. and French, Kenneth R.},
	Doi = {10.1016/0304-405X(93)90023-5},
	Journal = {Journal of Financial Economics},
	Number = 1,
	Pages = {3-56},
	Title = {Common risk factors in the returns on stocks and bonds},
	Url = {http://www.sciencedirect.com/science/article/pii/0304405X93900235},
	Volume = 33,
	Year = 1993,
	Bdsk-Url-1 = {http://www.sciencedirect.com/science/article/pii/0304405X93900235},
	Bdsk-Url-2 = {https://doi.org/10.1016/0304-405X(93)90023-5}}

@article{holthausen1992,
	Author = {Holthausen, Robert W. and Larcker, David F.},
	Doi = {https://doi.org/10.1016/0165-4101(92)90025-W},
	Journal = {Journal of Accounting and Economics},
	Pages = {373-411},
	Title = {The prediction of stock returns using financial statement information},
	Volume = 15,
	Year = 1992,
	Bdsk-Url-1 = {https://doi.org/10.1016/0165-4101(92)90025-W}}

@article{amihud1989,
	Author = {Amihud, Yakov and Mendelson, Haim},
	Doi = {https://doi.org/10.1111/j.1540-6261.1989.tb05067.x},
	Journal = {The Journal of Finance},
	Month = {June},
	Number = 2,
	Pages = {479-486},
	Title = {The Effects of Beta, Bid‐Ask Spread, Residual Risk, and Size on Stock Returns},
	Volume = 44,
	Year = 1989,
	Bdsk-Url-1 = {https://doi.org/10.1111/j.1540-6261.1989.tb05067.x}}

@article{ou1989,
	Author = {Ou, Jane A. and Penman, Stephen H.},
	Doi = {https://doi.org/10.1016/0165-4101(89)90017-7},
	Journal = {Journal of Accounting and Economics},
	Pages = {295-329},
	Title = {Financial Statement Analysis and the Prediction of Stock Returns},
	Volume = 11,
	Year = 1989,
	Bdsk-Url-1 = {https://doi.org/10.1016/0165-4101(89)90017-7}}

@article{bhandari1988,
	Author = {Bhandari, Laxmi Chand},
	Journal = {The Journal of Finance},
	Month = jun,
	Number = 2,
	Pages = {507-528},
	Title = {Debt/Equity Ratio and Expected Common Stock Returns: Empirical Evidence},
	Volume = 43,
	Year = 1988}

@article{debondt1985,
	Author = {De Bondt, Werner F. M. and Thaler, Richard},
	Doi = {https://doi.org/10.1111/j.1540-6261.1985.tb05004.x},
	Journal = {The Journal of Finance},
	Number = 3,
	Pages = {793-805},
	Title = {Does the Stock Market Overreact?},
	Volume = 40,
	Year = 1985,
	Bdsk-Url-1 = {https://doi.org/10.1111/j.1540-6261.1985.tb05004.x}}

@article{rendleman1982,
	Author = {Rendleman, Richard J. and Jones, Charles P. and Latan{\'e}, Henry A.},
	Doi = {https://doi.org/10.1016/0304-405X(82)90003-4},
	Journal = {Journal of Financial Economics},
	Month = nov,
	Number = 3,
	Pages = {269-287},
	Title = {Empirical anomalies based on unexpected earnings and the importance of risk adjustments},
	Url = {http://www.sciencedirect.com/science/article/B6VBX-45KNKNN-1V/1/89ba6a55ddcffb1d1116a5e3ae8300a8},
	Volume = 10,
	Year = 1982,
	Bdsk-Url-1 = {http://www.sciencedirect.com/science/article/B6VBX-45KNKNN-1V/1/89ba6a55ddcffb1d1116a5e3ae8300a8},
	Bdsk-Url-2 = {https://doi.org/10.1016/0304-405X(82)90003-4}}

@article{litzenberger1979,
	Author = {Litzenberger, Robert H. and Ramaswamy, Krishna},
	Doi = {https://doi.org/10.1016/0304-405X(79)90012-6},
	Journal = {Journal of Financial Economics},
	Month = jun,
	Number = 2,
	Pages = {163--195},
	Title = {The effect of personal taxes and dividends on capital asset prices : Theory and empirical evidence},
	Url = {http://www.sciencedirect.com/science/article/B6VBX-45KNKSN-3F/1/fa82aff26e3a14a28b5730fbd05750a0},
	Volume = 7,
	Year = 1979,
	Bdsk-Url-1 = {http://www.sciencedirect.com/science/article/B6VBX-45KNKSN-3F/1/fa82aff26e3a14a28b5730fbd05750a0},
	Bdsk-Url-2 = {https://doi.org/10.1016/0304-405X(79)90012-6}}

@article{basu1977,
	Author = {Basu, S.},
	Journal = {The Journal of Finance},
	Month = {June},
	Number = 3,
	Pages = {663-682},
	Title = {Investment Performance of Common Stocks in Relation to Their Price-Earnings Ratios: A Test of the Efficient Market Hypothesis},
	Volume = 32,
	Year = 1977}

@article{DT,
	Author = {Kent Daniel and Sheridan Titman},
	Journal = {Journal of Finance},
	Title = {Evidence on the Characteristics of Cross-Sectional Variation in Stock Returns},
	Year = {1997}}

@article{GFX,
	Author = {Giglio and Feng and Xiu},
	Journal = {Journal of Finance},
	Title = {Taming the Factor Zoo: a Test of New Factors},
	Year = {2020}}

@article{KNS1,
	Author = {Kozak and Nagel and Santosh},
	Journal = {Journal of Financial Economics},
	Title = {Shrinking the Cross Section},
	Year = {2019}}

@article{valentin,
	Author = {Haddad and Kozak and Santosh},
	Date-Modified = {2020-07-01 17:29:20 -0400},
	Journal = {Review of Financial Studies},
	Title = {Factor Timing},
	Year = {2020}}

@article{cohen,
	Author = {Cohen and Polk and Vuolteenhao},
	Journal = {Journal of Finance},
	Title = {The Value Spread},
	Year = {2003}}

@techreport{M2,
	Author = {Ehsani and Linnainmaa},
	Title = {Factor momentum and the momentum factor},
	Year = {2019}}

@article{KNS2,
	Author = {Kozak and Nagel and Santosh},
	Journal = {Journal of Finance},
	Title = {Interpreting Factor Models},
	Year = {Forthcoming}}

@article{...and,
	Author = {Harvey, Liu and Zhu},
	Journal = {Review of Financial Studies},
	Title = {... and the cross-section of Expected Returns},
	Year = {2016}}

@article{pontiff,
	Author = {McLean and Pontiff},
	Journal = {Journal of Finance},
	Title = {Does academic research destroy stock return predictability?},
	Year = {2016}}

@article{GHZ,
	Author = {Green, Hand and Zhang, 2017},
	Journal = {Review of Financial Studies},
	Title = {The Characteristics that Provide Independent Information about Average U.S. Monthly Stock Returns},
	Year = {2017}}

@article{HXZ,
	Author = {Hou, Xue and Zhang, 2017},
	Journal = {Review of Financial Studies},
	Title = {Replicating anomalies},
	Year = {2019}}

\newpage

\appendix
\setcounter{table}{0}
\setcounter{figure}{0}

\renewcommand\thetable{A.\arabic{table}}
\renewcommand\thefigure{B.\arabic{figure}}

\section{List of Factors}
\label{sec:list_factors}

\begin{table}[tbph]
\caption{List of Factor Names and References}
\label{tab:list_factor}
  \centering
  \begin{tabular}{ll}
    \toprule
    Factor                                           & Reference article\\
    \midrule
    Price earnings                                   & \cite{basu1977}\\
    Unexpected quarterly earnings                    & \cite{rendleman1982}\\
    Long term reversal                               & \cite{debondt1985}\\
    Debt equity                                      & \cite{bhandari1988}\\
    Change in inventory-to-assets                    & \cite{ou1989}\\
    Change in dividend per share                     & \cite{ou1989}\\
    Change in capital expenditures-to-assets         & \cite{ou1989}\\
    Return on total assets                           & \cite{ou1989}\\
    Debt repayment                                   & \cite{ou1989}\\
    Depreciation-to-PP\&E                            & \cite{holthausen1992}\\
    Change in depreciation-to-PP\&E                  & \cite{holthausen1992}\\
    Change in total assets                           & \cite{holthausen1992}\\
    Size                                             & \cite{fama1993}\\
    Book-to-market                                   & \cite{fama1993}\\
    Momentum                                         & \cite{jegadeesh1993}\\
    Sales growth                                     & \cite{lakonishok1994}\\
    Cash flow-to-price                               & \cite{lakonishok1994}\\
    Working capital accruals                         & \cite{Sloan1996}\\
    Sales-to-price                                   & \cite{barbee1996}\\
    Share turnover                                   & \cite{haugen1996}\\
    Cash flow-to-price variability                   & \cite{haugen1996}\\
    Trading volume trend                             & \cite{haugen1996}\\
    \bottomrule
    &\emph{\hspace{2cm} see next page}\\
  \end{tabular}
\end{table}

\newpage
\begin{table}[tbph]
\caption*{Table A.1 (Continued): List of Factor Names and References}
  \centering
  \begin{tabular}{ll}
    \toprule
    Factor                                           & Reference article\\
    \midrule
    Inventory                                        & \cite{abarbanell1998}\\
    Gross margin                                     & \cite{abarbanell1998}\\
    Capital expenditures                             & \cite{abarbanell1998}\\
    Industry momentum                                & \cite{moskowitz1999}\\
    F-score                                          & \cite{piotroski2000}\\
    Industry adjusted book-to-market                 & \cite{asness2000}\\
    Industry adjusted cash flow-to-price             & \cite{asness2000}\\
    Industry adjusted size                           & \cite{asness2000}\\
    Industry adjusted momentum                       & \cite{asness2000}\\
    Industry adjusted long term reversal             & \cite{asness2000}\\
    Industry adjusted short term reversal            & \cite{asness2000}\\
    Dollar volume                                    & \cite{chordia2001}\\
    Dollar volume coefficient of variation           & \cite{chordia2001}\\
    Share turnover coefficient of variation          & \cite{chordia2001}\\
    Change in inventory                              & \cite{thomas2002}\\
    Change in current assets                         & \cite{thomas2002}\\
    Depreciation                                     & \cite{thomas2002}\\
    Change in accounts receivable                    & \cite{thomas2002}\\
    Other accruals                                   & \cite{thomas2002}\\
    Illiquidity                                      & \cite{amihud2002}\\
    Liquidity                                        & \cite{pastor2003}\\
    Idiosyncratic return volatility x book-to-market & \cite{ali2003}\\
    Price x book-to-market                           & \cite{ali2003}\\
    \bottomrule
    &\emph{\hspace{2cm} see next page}\\
  \end{tabular}
\end{table}

\begin{table}[tbph]
\caption*{Table A.1 (Continued): List of Factor Names and References}
  \centering
  \begin{tabular}{ll}
    \toprule
    Factor                                           & Reference article\\
    \midrule
    Operating cash flow-to-price                     & \cite{desai2004}\\
    Abnormal corporate investment                    & \cite{titman2004}\\
    Accrual quality                                  & \cite{francis2004}\\
    Earnings persistence                             & \cite{francis2004}\\
    Smoothness                                       & \cite{francis2004}\\
    Value relevance                                  & \cite{francis2004}\\
    Timeliness                                       & \cite{francis2004}\\
    Tax income-to-book income                        & \cite{lev2004}\\
    Price delay                                      & \cite{hou2005}\\
    Firm age                                         & \cite{jiang2005}\\
    Duration                                         & \cite{jiang2005}\\
    Change in current operating assets               & \cite{richardson2005}\\
    Change in non-current operating liabilities      & \cite{richardson2005}\\
    Growth in capital expenditures                   & \cite{anderson2006}\\
    Growth in capital expenditures (alternative)     & \cite{anderson2006}\\
    Low volatility                                   & \cite{ang2006}\\
    Low beta $\Delta$VIX                             & \cite{ang2006}\\
    Zero trading days                                & \cite{liu2006}\\
    Composite issuance                               & \cite{daniel2006}\\
    Intangible return                                & \cite{daniel2006}\\
    Earnings surprises x revenue surprises           & \cite{jegadeesh2006}\\
    Industry concentration                           & \cite{hou2006}\\
    Change in shares outstanding                     & \cite{pontiff2008}\\
    Seasonality                                      & \cite{heston2008}\\
    Investment                                       & \cite{lyandres2008}\\
    Investment growth                                & \cite{xing2008}\\
    Change in asset turnover                         & \cite{soliman2008}\\
    \bottomrule
  \end{tabular}
\end{table}

\newpage
\section{Replication of Our Main Results Using Cross-Sectional Momentum}
  \label{sec:appendix_xs}

\begin{figure}[tbph]
  \centering
  \includegraphics[width=15cm]{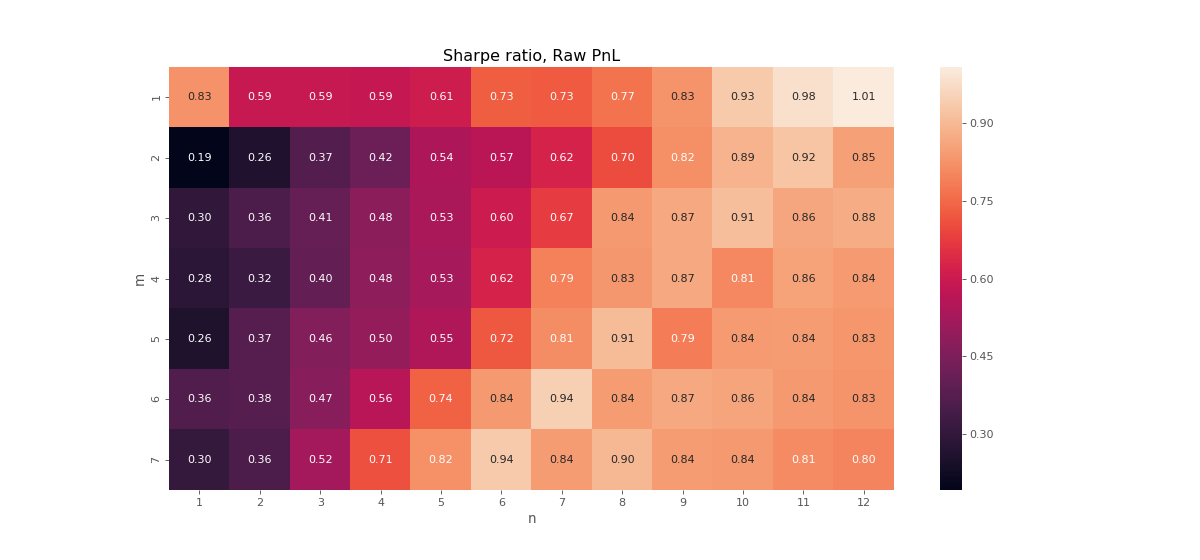}
  \caption{Sharpe ratio of the cross-sectional factor momentum. Factors are computed on CRSP 1,000 most liquid stocks (1963-2014) and are beta-neutralised and risk-managed.}
  \label{fig:sr_stock_factor_mom_xs}
\end{figure}

\begin{figure}[tbph]
  \centering
  \includegraphics[width=15cm]{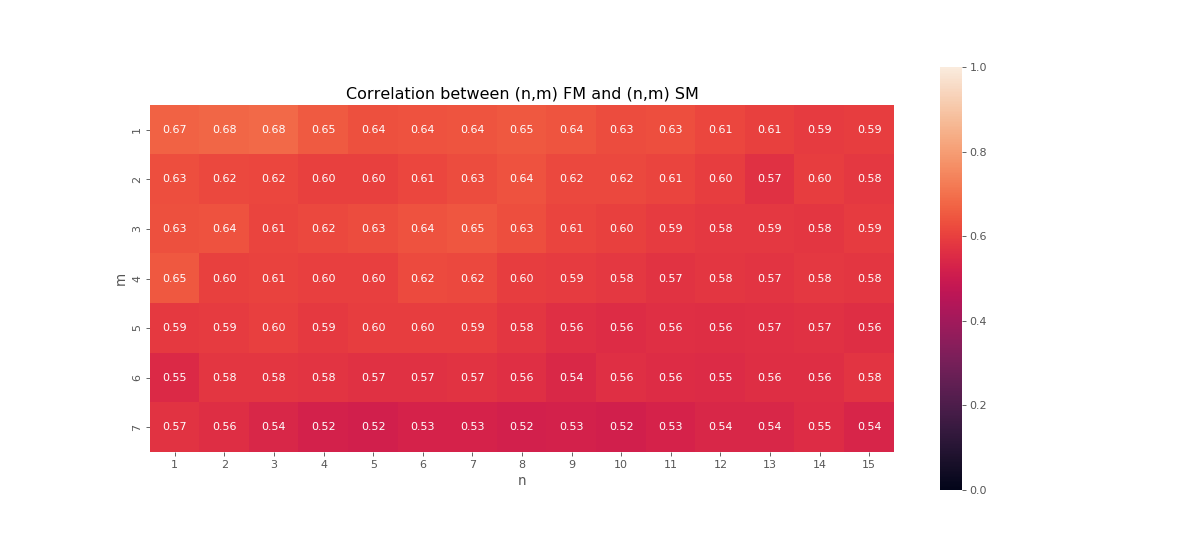}
  \caption{Monthly correlation of factor momentum (cross-sectional) and stock momentum.\\
  Stock returns are risk-managed and factors are beta-neutralised and risk-managed.
  The universe is CRSP 1,000 most liquid stocks (1963-2014).}
  \label{fig:corr_m_n_xs}
\end{figure}

\begin{figure}[tbph]
  \centering
  \includegraphics[width=15cm]{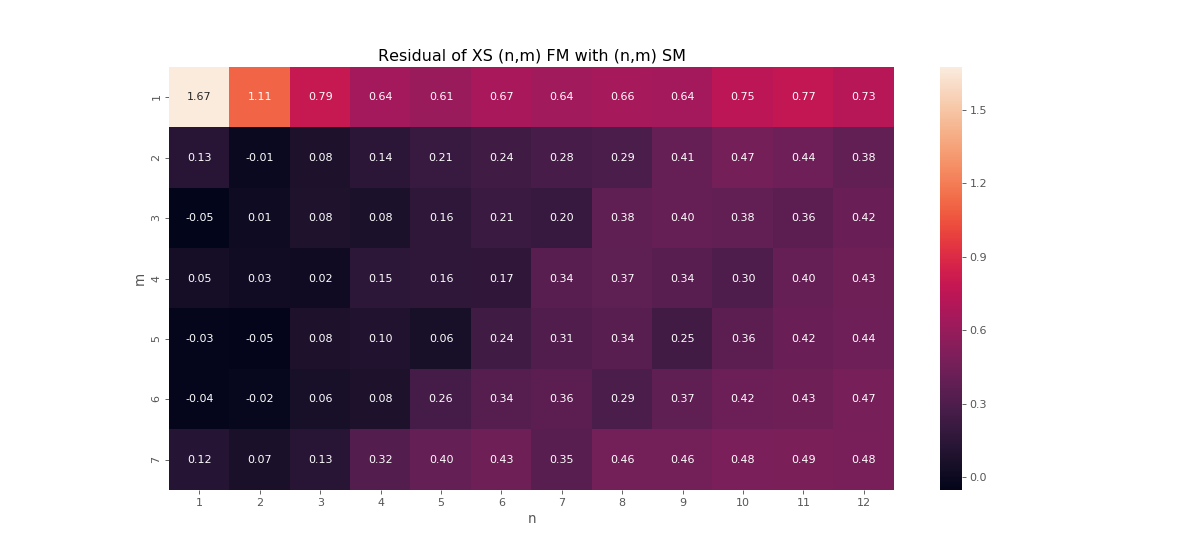}
  \caption{Sharpe ratio of the residual of factor momentum (cross-sectional)
  w.r.t. stock momentum.
  Stock returns are risk-managed and factors are beta-neutralised and risk-managed.
  The universe is CRSP 1,000 most liquid stocks (1963 to 2014).}
  \label{fig:corr_residual_xs}
\end{figure}

In this Appendix we replicate our main results from Figures \ref{fig:pnl_mom}, \ref{fig:sr_stock_factor_mom_ts} and \ref{fig:corr_m_n}  using cross-sectional factor momentum instead of time-series factor momentum. The results are essentially identical. Figure \ref{fig:sr_stock_factor_mom_xs} shows the PNL of the cross-sectional factor momentum for various values of $(m,n)$. Figure \ref{fig:corr_m_n_xs} reports the correlation between stock and (cross-sectional) factor momentum. Figure \ref{fig:corr_residual_xs} shows the Sharpe ratio of the residual of factor momentum on stock momentum.

\newpage

\section{Proofs}
\subsection{Covariance between Factor and Stock Momentum}
\label{proof: covariance}

Assume the vector of stock returns follows a single factor structure

$$r_t = \beta f_t + e_t$$

\noindent where we further assume that $e_t$ and $f_t$ are independent at all lags.

Then, the PNLs of (directional) factor momentum and stock momentum are given by the following two equations

\begin{align*}
\pi^F_{t+1} &= \underline{f}_t f_{t+1} \\
\pi^S_{t+1} &= \underline{r'}_t r_{t+1} 
\end{align*}

\noindent where the lower bar stands for any combination of lag $m$ and holding period $n$ until date $t$. 

To calculate the covariance of these two PNLs, we start with the expectation of the product

\begin{align*}
E(\pi^F_{t+1}\pi^S_{t+1}) &= E\left[E_{t}\left(\underline{f}_t f_{t+1} \underline{r'}_t r_{t+1}\right)\right] \\
&= E\left[\underline{f}_t \underline{r'}_t E_{t}\left(f_{t+1}  r_{t+1}\right)\right] \\
&= E\left[\underline{f}_t \underline{r'}_t E_{t}\left(f_{t+1} \left(\beta f_{t+1} + e_{t+1}\right)\right)\right]
\end{align*}

We note $\sigma_f^2$ and $\mu_t$ the conditional covariance and conditional expectation of $f_{t+1}$. We assume $f_t$ homoskedastic, so that the conditional variance is constant, but we allow the conditional mean to vary over time (there can be factor momentum for instance). We use the fact that $e_t$ and $f_t$ are independent so that

\begin{align*}
E(\pi^F_{t+1}\pi^S_{t+1}) = E\left[\underline{f}_t \underline{r'}_t \right]\beta\sigma^2_f + E\left[\underline{f}_t \underline{r'}_t \mu_t^2\right]\beta
\end{align*}

Finally, we exploit the fact that cumulative returns have the same factor structure as one period returns, i.e. that $\underline{r_t} = \beta \underline{f_t} + \underline{e_t}$. This leads to

\begin{align*}
E(\pi^F_{t+1}\pi^S_{t+1}) &= E\left[\underline{f}^2_t \right]\beta'\beta\sigma^2_f + E\left[\underline{f}^2_t \mu_t^2\right]\beta'\beta \\
&= \beta'\beta\left(\sigma^2_f E \underline{f}^2_t + E\left(\underline{f}_t \mu_t\right)^2 \right)
\end{align*}

Now, we need to compute the product of expectations

\begin{align*}
E(\pi^F_{t+1})E(\pi^S_{t+1}) &= \left( E\underline{f}_t \mu_t\right)^2 \beta'\beta \\
\end{align*}

Combining the two expressions yields:

\begin{align*}
\text{cov} \left(\pi^F_{t+1},\pi^S_{t+1}\right) &= \beta'\beta\left(\sigma^2_f E \underline{f}^2_t + E\left(\underline{f}_t^2 \mu_t^2\right) - \left( E\underline{f}_t \mu_t\right)^2 \right)  
\end{align*}

\noindent Noting that $\mu_t^2=-\sigma_f^2+E_t (f_{t+1}^2)$ we obtain
$$\text{cov} \left(\pi^F_{t+1},\pi^S_{t+1}\right) = \beta'\beta\text{var}\left(\underline{f}_t f_{t+1}\right) = \beta'\beta \text{var} \left(\pi^F_{t+1}\right) $$

\end{document}